\documentclass[english]{article}
\usepackage[T1]{fontenc}
\usepackage[latin9]{inputenc}
\usepackage{geometry}
\usepackage{color}
\usepackage{float}
\usepackage{amsmath}
\usepackage{amssymb} 
\usepackage{graphicx}

\makeatletter

\providecommand{\tabularnewline}{\\}

\PassOptionsToPackage{normalem}{ulem} 
\usepackage{babel}
\usepackage{cite}
\usepackage{color}
\usepackage{dsfont}
\usepackage{empheq}
\usepackage{framed}
\usepackage{mathrsfs}
\usepackage{needspace}
\usepackage{setspace}
\usepackage{tabularx}
\usepackage{tikz}
\usepackage{ulem}
\usepackage{url}

\let\@fnsymbol\@arabic

\newcommand{\mm}{\mu_{\mathrm{macro}}}
\newcommand{\lm}{\lambda_{\mathrm{macro}}}
\newcommand{\mh}{\mu_{\mathrm{micro}}}
\newcommand{\lh}{\lambda_{\mathrm{micro}}}
\newcommand{\me}{\mu_{e}}
\newcommand{\mc}{\mu_{c}}
\newcommand{\lle}{\lambda_{e}}

\newcommand{\mLc}{\me L_{c}^{2}}

\newcommand{\R}{\mathbb{R}}
\newcommand{\nablau}{\,\nabla u\,}
\newcommand{\p}{{P}}

\newcommand{\Curl}{\,\mathrm{Curl}}
\newcommand{\dev}{\, \mathrm{dev}}
\newcommand{\Div}{\mathrm{Div}}
\newcommand{\tr}{\, \mathrm{tr}}
\newcommand{\sym}{\, \mathrm{sym}\,}

\renewcommand{\skew}{\, \mathrm{skew}\,}

\renewcommand{\skew}{\, \mathrm{skew}}

\newcommand{\id}{\,\mathds{1}}
\newcommand{\vv}{v}

\definecolor{Green}{rgb}{0,0.52,0}

\makeatother

\usepackage{babel}
\begin{document}

\title{\vspace{-1.0cm}Modeling real phononic crystals via the weighted
relaxed micromorphic model with free and gradient micro-inertia}

\author{Angela Madeo\thanks{Angela Madeo, corresponding author, angela.madeo@insa-lyon.fr, LGCIE
SMS-ID, INSA-Lyon, Universit\'e de Lyon, 20 avenue Albert Einstein,
69621, Villeurbanne cedex and IUF, Institut Universitaire de France,
1 rue Descartes, 75231 Paris Cedex 05, France}, Manuel Collet\thanks{Manuel Collet, LTDS UMR-CNRS 5513, Ecole Centrale de Lyon, 36 avenue
Guy de Collongue, 69134 Ecully, France}, Marco Miniaci\thanks{Marco Miniaci, Universit\'e du Havre, Laboratoire Ondes et Milieux Complexes,
UMR CNRS 6294, 75 Rue Bellot, 76600 Le Havre, France}, K\'evin Billon\thanks{K\'evin Billon: FEMTO-ST, Applied Mechanics Department, UMR-CNRS 6174,
24 chemin de l\textquoteright \'epitaphe, 25000 Besancon, France}, Morvan Ouisse\thanks{Morvan Ouisse, FEMTO-ST, Applied Mechanics Department, UMR-CNRS 6174,
24 chemin de l\textquoteright \'epitaphe, 25000 Besancon, France}, \\and Patrizio Neff\thanks{Patrizio Neff, patrizio.neff@uni-due.de, Head of Chair for 	Nonlinear Analysis and Modelling, Fakultät für Mathematik, Universität Duisburg-Essen,  Mathematik-Carrée, Thea-Leymann-Straße 9, 45127 Essen, Germany}}

\maketitle
\addtocounter{footnote}{5} 
\begin{abstract}
In this paper the relaxed micromorphic continuum model with weighted free and gradient micro-inertia is used to describe the dynamical behavior of a real two-dimensional phononic crystal for a wide range of wavelengths. In particular, a periodic structure with specific micro-structural topology and mechanical properties, capable of opening a phononic band-gap, is chosen with the criterion of showing a low degree of anisotropy (the band-gap is almost independent of the direction of propagation of the traveling wave). A Bloch wave analysis is performed to obtain the dispersion curves and the corresponding vibrational modes of the periodic structure. A linear-elastic, isotropic, relaxed micromorphic model including both a free micro-inertia (related to free vibrations of the microstructures) and a gradient micro-inertia (related to the motions of the microstructure which are coupled to the macro-deformation of the unit cell) is introduced and particularized to the case of plane wave propagation. The parameters of the relaxed model, which are independent of frequency, are then calibrated on the dispersion curves of the phononic crystal showing an excellent agreement in terms of both dispersion curves and vibrational modes. Almost all the homogenized elastic parameters of the relaxed micromorphic model result to be determined. This opens the way to the design of morphologically complex meta-structures which make use of the chosen phononic structure as the basic building block and which preserve its ability of ``stopping'' elastic wave propagation at the scale of the structure.\vspace{0.5cm}
\end{abstract}

	\vspace{0.5cm}
\hspace{-0.55cm}\textbf{AMS 2010 subject classification}: 74A10 (stress),
74A30 (non-simple materials), 74A60 (micro-mechanical theories), 74E15
(crystalline structure), 74M25 (micro-mechanics), 74Q15 (effective
constitutive equations)


\newpage\tableofcontents{}\vspace{1.2cm}

\section{Introduction}

A new class of innovative engineered materials, also known as phononic/photonic
crystals or metamaterials, showing exotic behaviours with respect
to both mechanical and electromagnetic wave propagation are recently
attracting great interest for their unique unconventional properties,
such as frequency band gaps \cite{armenise2010phononic,martinez1995sound,yablonovitch1993photonic},
negative refraction \cite{morvan2010experimental,pendry2000negative},
cloaking \cite{colquitt2014transformation,pendry2006controlling},
filtering/focusing capabilities \cite{brun2010dynamics}, etc. Among
these unorthodox properties, the ability to \textquotedblleft stop\textquotedblright{}
or \textquotedblleft bend\textquotedblright{} the propagation of waves
of light or sound with no energetic inputs is the main aspect disclosing
rapid and previously unimaginable technological advancements.

Specifically, in the electromagnetic domain, these metamaterials might
be used for rendering aircrafts or other vehicles undetectable to
radar, for making objects invisible to the human eye \cite{pendry2006controlling},
or to realize the so-called super-lenses that would allow the human
eye to see single viruses or nano-organisms \cite{craster2013acoustic}.
Other important applications concern the revolution of the electronics
in communication and information management systems, using light instead
of electrons as the information carrier, through photonic band-gap
materials \cite{blanco2000large}. Finally, in order to obtain wider
and frequency tunable band-gaps several approaches have been proposed
by varying the properties of the materials as well as their spatial
ordered/disordered arrangement \cite{florescu2009designer,florescu2009complete,haberko2013fabrication,man2013photonic,lin1999three,yamamoto1999fabrication}.

Notwithstanding the interest raised by such \textquotedblleft electromagnetic
metamaterials\textquotedblright , they will not make the object of
the present paper which will be instead centered on the study of \textquotedblleft mechanical
metamaterials\textquotedblright . 

As far as phononic crystals and mechanical metamaterials are concerned,
regrettably, their introduction into the current technology appears
less advanced, although many potential applications of practical implementation
have already been proposed: from seismic protection \cite{colombi2016seismic,miniaci2016large}
to environmental noise reduction\cite{martinez1995sound,morandi2016standardised},
from sub-wavelength imaging to focusing\cite{bigoni2013elastic},
acoustic cloaking \cite{misseroni2016cymatics} and even thermal control\cite{maldovan2013sound}.

Such materials also pushed innovative naval, automotive and aeronautical
vehicles conception in view of absorbing external solicitations and
shocks, thereby drastically improving their internal comfort. What's
more, civil engineering structures which are built in the vicinity
of sources of vibrations such as metro lines, tramways, train stations,
and so forth would take advantage of the use of these metamaterials
to ameliorate the enjoyment of internal and external environments
\cite{huang2013attenuation}.

Based on the same principle, passive engineering devices perfectly
able to insulate from noise \cite{jimenez2016ultra} could be easily
conceived and produced at relatively low costs. The conception of
waveguides used to optimize energy transfers by collecting waves in
slabs \cite{wang2013multiple} or wires, as well as the design of
wave screens employed to protect from any sort of mechanical wave
could also see a new technological revolution. And many other unprecedented
applications that, at this juncture, we cannot even envision could
be abruptly disclosed once such metamaterials would become easily
accessible.

\medskip{}

In this paper, we focus our attention on those metamaterials which
are able to \textquotedblleft stop\textquotedblright{} wave propagation,
i.e. metamaterials in which waves within precise frequency ranges
cannot propagate. Such frequency intervals at which wave inhibition
occurs are known as frequency band-gaps and their intrinsic characteristics
(characteristic values of the gap frequency, extension of the band-gap,
etc.) strongly depend on the metamaterial microstructure. 

The approach that we use here to describe the mechanical behavior
of band-gap metamaterials is in complete rupture to the classical
approaches that are nowadays used to study phononic crystals. Indeed,
the most spread approach to the modeling of band-gap metamaterials
is that of starting from a precise microstructure and to derive the
dispersive properties of the equivalent medium using upscaling arguments
or numerical homogenization techniques (see e.g. \cite{liu2000locally,pham2013transient,sridhar2016homogenization,spadoni2009phononic,aurialt2012long,boutin2003homogenisation,boutin2010generalized,boutin2011generalized}).
If such methods allow, on the one hand, to make a direct comparison
between the micro and the macro properties, on the other hand, they
are often strongly limited to large wavelengths. 

From a different perspective, we start from a macroscopic (directly
defined at the homogenized level) linear-elastic relaxed micromorphic
model (which is known to be able to describe band-gap behaviors and
to be well-posed \cite{dagostino2016panorama,madeo2014band,madeo2015wave,madeo2016complete,madeo2016first,madeo2016reflection,neff2015relaxed,neff2014unifying})
and we determine the parameters of our model on real metamaterials
by an inverse approach. Similarly to classical isotropic linear-elasticity
where two parameters (Young modulus and Poisson ratio) are needed
to describe the average behavior of a large class of engineering materials,
in the same way in isotropic, linear-elastic enriched elasticity few
extra parameters will be needed to describe the averaged behavior
of a relatively huge class of metamaterials sharing the common characteristic
property of stopping wave propagation. Such parameters, being constitutive
parameters, do not depend on frequency as it is instead usually the
case when dealing with homogenized media arising from classical homogenization
techniques. 

The fact of determining the coefficients of an enriched continuum
model on real band-gap metamaterials with given microstructure is
of paramount importance for the subsequent implementation of the model
in Finite Element codes and for an effective design of morphologically
complex band-gap metastructures.

\subsection{Notations}

In this contribution, we denote by $\R^{3\times3}$ the set of real
$3\times3$ second order tensors, written with capital letters. We
denote respectively by $\cdot\:$, $:$ and $\left.\langle\cdot,\cdot\right.\rangle$
a simple and double contraction and the scalar product between two
tensors of any suitable order\footnote{For example, $(A\cdot v)_{i}=A_{ij}v_{j}$, $(A\cdot B)_{ik}=A_{ij}B_{jk}$,
$A:B=A_{ij}B_{ji}$, $(C\cdot B)_{ijk}=C_{ijp}B_{pk}$, $(C:B)_{i}=C_{ijp}B_{pj}$,
$\left.\langle v,w\right.\rangle=v\cdot w=v_{i}w_{i}$, $\left.\langle A,B\right.\rangle=A_{ij}B_{ij}$
etc.}. Everywhere we adopt the Einstein convention of sum over repeated
indices if not differently specified. The standard Euclidean scalar
product on $\R^{3\times3}$ is given by $\langle{X},{Y}\rangle_{\R^{3\times3}}=\tr({X\cdot Y^{T}})$,
and thus the Frobenius tensor norm is $\|{X}\|^{2}=\langle{X},{X}\rangle_{\R^{3\times3}}$.
In the following we omit the index $\R^{3},\R^{3\times3}$ if no confusion
can arise. The identity tensor on $\R^{3\times3}$ will be denoted
by $\id$, so that $\tr({X})=\langle{X},{\id}\rangle$. 

\medskip{}

We consider a body which occupies a bounded open set $B$ of the three-dimensional
Euclidian space $\R^{3}$ and assume that its boundary $\partial B$
is a smooth surface of class $C^{2}$. An elastic material fills the
domain $B\subset\R^{3}$ and we refer the motion of the body to rectangular
axes $Ox_{i}$. 

For vector fields $v$ with components in ${\rm H}^{1}(B)$, i.e.
$v=\left(v_{1},v_{2},v_{3}\right)^{T}\,,v_{i}\in{\rm H}^{1}(B),$
we define \break $\nabla\,v=\left((\nabla\,v_{1})^{T},(\nabla\,v_{2})^{T},(\nabla\,v_{3})^{T}\right)^{T}$,
while for tensor fields $P$ with rows in ${\rm H}({\rm curl}\,;B)$,
resp. ${\rm H}({\rm div}\,;B)$, i.e. \break $P=\left(P_{1}^{T},P_{2}^{T},P_{3}^{T}\right)$,
$P_{i}\in{\rm H}({\rm curl}\,;B)$ resp. $P_{i}\in{\rm H}({\rm div}\,;B)$
we define ${\rm Curl}\,P=\left(({\rm curl}\,P_{1})^{T},({\rm curl}\,P_{2})^{T},({\rm curl}\,P_{3})^{T}\right)^{T},$
${\rm Div}\,P=\left({\rm div}\,P_{1},{\rm div}\,P_{2},{\rm div}\,P_{3}\right)^{T}.$ 

\vspace{0.2cm}

As for the kinematics of the considered micromorphic continua, we
introduce the functions
\[
\chi(X,t):B\subset\mathbb{R}^{3}\rightarrow\mathbb{R}^{3},\qquad\qquad P(X,t):B\subset\mathbb{R}^{3}\rightarrow\mathbb{R}^{3\times3},
\]
which are known as \textit{placement} vector field and \textit{micro-distortion}
tensor, respectively. The physical meaning of the placement field
is that of locating, at any instant $t$, the current position of
the material particle $X\in B$, while the micro-distortion field
describes deformations of the microstructure embedded in the material
particle $X$. As it is usual in continuum mechanics, the displacement
field can also be introduced as the function $u(X,t):B\subset\mathbb{R}^{3}\rightarrow\mathbb{R}^{3}$
defined as
\[
u(X,t)=\chi(X,t)-X.
\]

\section{The relaxed micromorphic model with weighted free and gradient micro
inertia}

In recent previous contributions \cite{dagostino2016panorama,ghiba2014relaxed,madeo2014band,madeo2015wave,madeo2016complete,neff2015relaxed,neff2014unifying},
the relaxed micromorphic model has been introduced as that enriched
model of the micromorphic type which, in the linear-elastic, isotropic
case, features a strain energy density of the form
\begin{align}
W= & \underbrace{\me\,\lVert\sym\left(\nablau-\p\right)\rVert^{2}+\frac{\lle}{2}\left(\mathrm{tr}\left(\nablau-\p\right)\right)^{2}}_{\mathrm{{\textstyle isotropic\ elastic-energy}}}+\hspace{-0.1cm}\underbrace{\mc\,\lVert\skew\left(\nablau-\p\right)\rVert^{2}}_{\mathrm{{\textstyle rotational\ elastic\ coupling}}}\hspace{-0.1cm}\label{eq:Ener-General}\\
 & \quad+\underbrace{\mh\,\lVert\sym\p\rVert^{2}+\frac{\lh}{2}\,\left(\mathrm{tr}\p\right)^{2}}_{\mathrm{{\textstyle micro-self-energy}}}+\underbrace{\frac{\mLc}{2}\,\lVert\Curl\p\rVert^{2}}_{\mathrm{{\textstyle isotropic\ curvature}}}\,,\nonumber 
\end{align}
where all the introduced constitutive coefficients are positive constant
(the Cosserat couple modulus $\mu_{c}$ can also be vanishing in some
special cases without affecting the well-posedness of the model \textcolor{black}{(see
\cite{neff2006cosserat})}. This particular constitutive form of the
strain energy density will be used in the present paper as descriptive
of a certain class of metamaterials with particular topologies that
will be shown to exhibit band-gap behaviors\footnote{This same energy could be used to describe, from a macroscopic point
of view, the behavior of band-gap metamaterials obtained using piezoelectric
patches, as those presented e.g. in \cite{Yi2016flexural}. }. Such constitutive choice is dictated by the previous works on this
subject \cite{madeo2014band,madeo2015wave,madeo2016complete,madeo2016reflection,madeo2016role}
showing that the relaxed micromorphic model is the most effective
enriched continuum model that can be used for the simultaneous description
of band-gaps and non-localities in metamaterials. As we will show
in the second part of the paper, the fitting of the parameters on
the basis of the dispersion curves, as presented in the present work
allows to have a first estimate of the elastic parameters of the relaxed
micromorphic model. On the other hand, a fitting procedure based on
the dispersion curves alone is not precise enough to allow an accurate
estimation of the characteristic length of the considered metamaterial.
In this paper, only the main elastic parameters of the model will
be then determined, while the effect and associated estimate of the
characteristic length will be analyzed in further works where the
fitting will be based on the use of the transmission coefficient as
done already in \cite{madeo2016first}.

\medskip{}

As far as the adopted kinetic energy is concerned, we consider (as
done in \cite{dagostino2016panorama}) a Cartan-Lie decomposition
of the free micro-inertia $\left\Vert \p_{,t}\right\Vert ^{2}$, as
well as of the gradient micro-inertia $\left\Vert \nablau_{,t}\right\Vert ^{2}$
(as presented in \cite{madeo2016role}). The kinetic energy density
that we thus retain in this paper to model the mechanical behavior
of the targeted real phononic crystals takes the following form:

\begin{gather}
J=\hspace{-0.1cm}\underbrace{\frac{1}{2}\rho\left\Vert u_{,t}\right\Vert ^{2}}_{\text{Cauchy inertia}}+\hspace{0.1cm}\underbrace{\frac{1}{2}\eta_{1}\left\Vert \dev\sym\p_{,t}\right\Vert ^{2}+\frac{1}{2}\eta_{2}\left\Vert \skew\p_{,t}\right\Vert ^{2}+\frac{1}{6}\eta_{3}\tr\left(\p_{,t}\right)^{2}}_{\text{weighted free micro-inertia}},\nonumber \\
\label{eq:Kinetic}\\
+\hspace{0.1cm}\underbrace{\frac{1}{2}\overline{\eta}_{1}\left\Vert \dev\sym\nablau_{,t}\right\Vert ^{2}+\frac{1}{2}\overline{\eta}_{2}\left\Vert \skew\nablau_{,t}\right\Vert ^{2}+\frac{1}{6}\overline{\eta}_{3}\tr\left(\nablau_{,t}\right)^{2}}_{\text{weighted gradient micro-inertia}},\nonumber 
\end{gather}
We discuss here briefly the inertia terms appearing in Eq. \eqref{eq:Kinetic}
in order to give an explanation of the adopted nomenclature as well
as a general interpretation of the respective physical meaning associated
to each term:
\begin{itemize}
\item The Cauchy inertia term $\frac{1}{2}\rho\left\Vert u_{,t}\right\Vert ^{2}$
is the macroscopic inertia introduced in classical linear elasticity.
It allows to describe the vibrations associated to the macroscopic
displacement field. In an enriched continuum mechanical modeling framework,
this means that such terms account for the inertia to vibration of
the unit cells considered as material points (or Representative Volume
Elements) with apparent mass density $\rho$.
\item The term $\frac{1}{2}\eta\left\Vert \p_{,t}\right\Vert ^{2}$ accounts
for the inertia of the microstructure alone: we called $\eta$ \textit{free
micro-inertia} \cite{madeo2016role} since it represents the inertia
of the microstructure seen as a micro-system whose vibration can be
independent of the vibration of the unit cells. An inertia term of
this type is mandatory whenever one considers an enriched model of
the micromorphic type, i.e. a model that features an enriched kinematics
$(u,P)$. Indeed, it would be senseless to introduce an enriched kinematics,
an enriched constitutive form for the strain energy density and then
avoid to introduce this free micro-inertia in the model. It would
be like introducing a complex constitutive structure to describe in
detail the mechanical behavior of microstructured materials while
not giving to the model the possibility of activating the vibrations
of such microstructures. The free micro-inertia allows us to account
for the vibrations of the microstructures that typically appear for
high frequencies (i.e. small wavelengths comparable with the characteristic
size of the microstructure) in a huge variety of mechanical metamaterials.
As we will show in more detail in the remainder of this paper, the
Cartan-Lie decomposition of the tensor $P_{,t}$ in its $\dev\sym$
(trace-free symmetric-), $\skew$ (skew-symmetric-) and $\tr$ (trace-)
part allows for the independent control of the cut-off frequencies
of the optic branches. This feature will be crucial for the fitting
of the parameters of our model on the real metamaterials targeted
in this paper.
\item The \textit{gradient micro-inertia} term is of the type $\bar{\eta}\left\Vert \nablau_{,t}\right\Vert ^{2}$
and, when split using a Cartan-Lie decomposition, it takes the form
shown in Eq. \eqref{eq:Kinetic} (see also \cite{madeo2016role}).
Such term allows to account for some specific vibrations of the microstructure
which are directly coupled to the deformation of the unit cell at
the macro scale. In other words, this term allows to account for the
inertia of the motions of the microstructure which are generated as
a consequence of the deformation of the unit cell as a whole. This
gradient micro-inertia term brings additional informations with respect
to the free micro-inertia term previously described and this fact
is translated on the behavior of some dispersion curves that, as we
will see, can be flattened when increasing the value of $\bar{\eta}_{1}$,
$\bar{\eta}_{2}$ or $\bar{\eta}_{3}$.
\end{itemize}
\medskip{}
The action functional $\mathcal{A}$ of the considered model can be
introduced as 
\begin{equation}
\mathcal{A}=\int_{0}^{T}\int_{B_{L}}(J-W)\,dv\,dt,\label{eq:Action}
\end{equation}
where $[0,T]$ is the interval of time during which the motion of
the considered micromorphic system wants to be observed. Following
standard variational arguments, the equations of motion of the system
can be obtained by making the action functional \eqref{eq:Action}
stationary and take the form (see also \cite{ghiba2014relaxed,neff2015relaxed,neff2014unifying,madeo2014band,madeo2015wave})

\begin{align}
\rho\,u_{,tt}-\Div[\,\mathcal{I}\,] & =\Div\left[\,2\,\me\,\sym\left(\nablau-\p\right)+\lle\,\tr\left(\nablau-\p\right)\id+2\,\mc\,\skew\left(\nablau-\p\right)\,\right],\nonumber \\
\eta_{1}\,\dev\sym\,\p_{,tt} & =2\,\mu_{e}\,\dev\sym\,\left(\nabla u-P\right)-2\,\mh\dev\sym\,P-\mLc\,\dev\sym\left(\Curl\,\Curl\,P\right),\nonumber \\
\eta_{2}\skew\,\p_{,tt} & =2\,\mu_{c}\skew\,\left(\nabla u-P\right)-\mLc\,\skew\left(\Curl\,\Curl\,P\right)\nonumber \\
\frac{1}{3}\eta_{3}\tr\p_{,tt} & =\left(\frac{2}{3}\mu_{e}+\lambda_{e}\right)\tr\left(\nabla u-P\right)-\left(\frac{2}{3}\mh+\lh\right)\tr\p-\frac{1}{3}\mLc\tr\left(\Curl\,\Curl\,P\right).\label{eq:Dyn}
\end{align}
where, for compactness, we set $\mathcal{I}=\overline{\eta}_{1}\,\dev\sym\nablau_{,tt}+\overline{\eta}_{2}\,\skew\nablau_{,tt}+\frac{1}{3}\overline{\eta}_{3}\tr\left(\nablau_{,tt}\right)$.

\subsection{Plane wave ansatz}

We rapidly recall in this subsection how, starting from the equations
of motion in strong form for the relaxed micromorphic medium, it is
possible to obtain the associated dispersion curves by following standard
techniques. We start by making a plane-wave ansatz which means that
we assume that all the 12 scalar components of the unknown fields\footnote{In what follows, we will not differentiate anymore the Lagrangian
space variable $X$ and the Eulerian one $x$. In general, such undifferentiated
space variable will be denoted as $x=(x_{1},x_{2},x_{3})^{T}$. } $u(x,t)$ and $P(x,t)$ only depend on the component $x_{1}$ of
the space variable $x$ which is also assumed to be the direction
of the traveling wave. With this unique assumption, together with
the introduction of the new variables 
\begin{align}
\p^{S} & :=\frac{1}{3}\tr\left(\p\right), & \p_{\left[ij\right]} & :=\left(\skew\p\right)_{ij}=\frac{1}{2}\left(\p_{ij}-\p_{ji}\right),\label{Decom}\\
\p^{D} & :=\p_{11}-\p^{S}, & \p_{\left(ij\right)} & :=\left(\sym\p\right)_{ij}=\frac{1}{2}\left(\p_{ij}+\p_{ji}\right),\nonumber \\
P^{V} & :=P_{22}-P_{33}, &  & i,j=\{1,2,3\},\nonumber 
\end{align}
the equations of motions \eqref{eq:Dyn} can be simplified and rewritten,
after suitable manipulation, as (see \cite{dagostino2016panorama,madeo2014band,madeo2015wave,madeo2016complete}
for additional details):
\begin{itemize}
\item a set of three equations only involving longitudinal quantities: 
\begin{align}
\rho\,\ddot{u}_{1}-\frac{2\,\overline{\eta}_{1}+\overline{\eta}_{3}}{3}\,\ddot{u}_{1,11} & =\left(2\,\me+\lle\right)u_{1,11}-2\me\,P_{,1}^{D}-(2\mu_{e}+3\lambda_{e})\,P_{,1}^{S}\,,\vspace{0.4cm}\nonumber \\
\eta_{1}\,\ddot{P}^{D} & =\frac{4}{3}\,\me\,u_{1,1}+\frac{1}{3}\,\mLc\,P_{,11}^{D}-\frac{2}{3}\,\mLc P_{,11}^{S}-2\left(\me+\mh\right)\,P^{D}\,,\vspace{0.4cm}\label{Long}\\
\eta_{3}\,\ddot{P}^{S} & =\frac{2\,\me+3\,\lle}{3}\,u_{1,1}-\frac{1}{3}\,\mLc P_{,11}^{D}+\frac{2}{3}\,\mLc P_{,11}^{S}\nonumber \\
 & \quad-\left(2\,\me+3\,\lle+2\,\mh+3\,\lh\right)\,P^{S}\,,\nonumber 
\end{align}
\item two sets of three equations only involving transverse quantities in
the $\xi$-th direction, with $\xi=2,3$: 
\begin{align}
\rho\,\ddot{u}_{\xi}-\frac{\,\overline{\eta}_{1}+\overline{\eta}_{2}}{2}\,\ddot{u}_{\xi,11} & =\left(\me+\mc\right)u_{\xi,11}-2\,\me\,P_{\left(1\xi\right),1}+2\,\mc\,P_{\left[1\xi\right],1},\vspace{0.4cm}\nonumber \\
\eta_{1}\,\ddot{P}_{\left(1\xi\right)} & =\me\,u_{\xi,1}+\frac{1}{2}\,\mLc\,P_{(1\xi)}{}_{,11}+\frac{1}{2}\,\mLc\,P_{\left[1\xi\right],11}\label{Trans}\\
 & \quad-2\left(\me+\mh\right)\,P_{(1\xi)},\vspace{0.4cm}\nonumber \\
\eta_{2}\,\ddot{P}_{\left[1\xi\right]} & =-\mc\,u_{\xi,1}+\frac{1}{2}\,\mLc\,P_{(1\xi),11}+\frac{1}{2}\,\mLc P_{\left[1\xi\right]}{}_{,11}-2\,\mc\,P_{\left[1\xi\right]},\nonumber 
\end{align}
\end{itemize}

\begin{itemize}
\item one equation only involving the variable $P_{\left(23\right)}$: 
\begin{align}
\eta_{1}\,\ddot{P}_{\left(23\right)}=-2\left(\me+\mh\right)P_{\left(23\right)}+\mLc P_{\left(23\right),11},\label{Shear}
\end{align}
\item one equation only involving the variable $P_{\left[23\right]}$ :
\[
\eta_{2}\,\ddot{P}_{\left[23\right]}=-2\,\mc\,P_{\left[23\right]}+\mLc P_{\left[23\right],11},
\]
\item one equation only involving the variable $P^{V}$: 
\begin{align}
\eta_{1}\,\ddot{P}^{V}=-2\left(\me+\mh\right)P^{V}+\mLc P_{,11}^{V}.\label{VolumeVariation}
\end{align}
\end{itemize}
Once that this simplified form of the equations of motion is obtained,
we look for a wave form solution of the type
\begin{alignat}{2}
\vv_{1}=\beta_{1}\,e^{i(kx_{1}-\omega t)},\qquad & \vv_{2}=\beta_{2}\,e^{i(kx_{1}-\omega t)},\qquad & \vv_{3}=\beta_{3}\,e^{i(kx_{1}-\omega t)},\nonumber \\
\label{eq:WaveForm}\\
\vv_{4}=\beta_{4}\,e^{i(kx_{1}-\omega t)},\qquad & \vv_{5}=\beta_{5}\,e^{i(kx_{1}-\omega t)},\qquad & \vv_{6}=\beta_{6}\,e^{i(kx_{1}-\omega t)},\nonumber 
\end{alignat}
where $\beta_{1}$, $\beta_{2}$, $\beta_{3}\in\mathbb{C}^{3}$ and
$\beta_{4}$, $\beta_{5}$, $\beta_{6}\in\mathbb{C}$ are the unknown
amplitudes of the considered waves, $\omega$ is the frequency, $k$
the wavenumber and where, for compactness of notation, we set

\begin{gather}
\vv_{1}:=\left(u_{1},P_{11}^{D},P^{S}\right),\qquad\vv_{2}:=\left(u_{2},P_{(12)},P_{[12]}\right),\qquad\vv_{3}:=\left(u_{3},P_{(13)},P_{[13]}\right),\nonumber \\
\label{eq:Unknowns}\\
\vv_{4}:=P_{\left(23\right)},\qquad\vv_{5}:=P_{[23]},\qquad\vv_{6}:=P^{V}.\nonumber 
\end{gather}
Replacing the wave-form \eqref{eq:WaveForm}-\eqref{eq:Unknowns}
in the equations of motion \eqref{Long}-\eqref{VolumeVariation}
and simplifying, we end up with the following systems of algebraic
equations
\begin{equation}
A_{1}(\omega,k)\cdot\beta_{1}=0,\qquad\qquad A_{\tau}(\omega,k)\cdot\beta_{\tau}=0,\quad\tau=2,3,\qquad\qquad A_{4}(\omega,k)\cdot\alpha=0,\label{AlgSys}
\end{equation}
where we set $\alpha=\left(\beta_{4},\beta_{5},\beta_{6}\right)$
and

\vspace{0.2cm}

\begin{align}
A_{1}(\omega,k)\, & =\left(\begin{array}{ccc}
-\omega^{2}\left(1+k^{2}\,\frac{2\,\overline{\eta}_{1}+\overline{\eta}_{3}}{3\,\rho}\right)+c_{p}^{2}\,k^{2} & \,i\:k\,2\,\me/\rho\  & i\:k\:\left(2\,\me+3\,\lle\right)/\rho\\
\\
-i\:k\,\frac{4}{3}\,\me/\eta_{1} & -\omega^{2}+\frac{1}{3}k^{2}c_{m1}^{2}+\omega_{s}^{2} & -\frac{2}{3}\,k^{2}c_{m1}^{2}\\
\\
-\frac{1}{3}\,i\,k\:\left(2\,\me+3\,\lle\right)/\eta_{3} & -\frac{1}{3}\,k^{2}\,c_{m3}^{2} & -\omega^{2}+\frac{2}{3}\,k^{2}\,c_{m3}^{2}+\omega_{p}^{2}
\end{array}\right),\nonumber \\
\nonumber \\
A_{2}(\omega,k)=A_{3}(\omega,k)\, & =\left(\begin{array}{ccc}
-\omega^{2}\left(1+k^{2}\,\frac{\overline{\eta}_{1}+\overline{\eta}_{2}}{2\,\rho}\right)+k^{2}c_{s}^{2}\  & \,i\,k\,2\,\me/\rho\  & -i\,\frac{\eta_{2}}{\rho}\,\omega_{r}^{2}k\,,\\
\\
-\,i\,k\,\me/\eta_{1}, & -\omega^{2}+\frac{1}{2}\,c_{m1}^{2}\,k^{2}+\omega_{s}^{2} & \frac{1}{2}\,c_{m1}^{2}\,k^{2}\\
\\
\frac{1}{2}\,i\,\omega_{r}^{2}\,k & \frac{1}{2}\,c_{m2}^{2}\,k^{2} & -\omega^{2}+\frac{1\,}{2}c_{m2}^{2}\,k^{2}+\omega_{r}^{2}
\end{array}\right),\nonumber \\
\\
A_{4}(\omega,k)\, & =\left(\begin{array}{ccc}
-\omega^{2}+c_{m1}^{2}\,k^{2}+\omega_{s}^{2} & 0 & 0\\
\\
0 & -\omega^{2}+c_{m2}^{2}\,k^{2}+\omega_{r}^{2} & 0\\
\\
0 & 0 & -\omega^{2}+c_{m1}^{2}\,k^{2}+\omega_{s}^{2}
\end{array}\right).\nonumber 
\end{align}

In the definition of the matrices $A_{i}$, $i=\{1,2,3,4\}$ the following
characteristic quantities have also been introduced:

\begin{align}
\omega_{s} & =\sqrt{\frac{2\,(\mu_{e}+\mh)}{\eta_{1}}},\qquad\omega_{r}=\sqrt{\frac{2\,\mu_{c}}{\eta_{2}}},\qquad\omega_{p}=\sqrt{\frac{(3\,\lambda_{e}+2\,\mu_{e})+(3\,\lh+2\,\mh)}{\eta_{3}}},\vspace{1.2mm}\nonumber \\
\label{eq:characteristic_quantities}\\
c_{m1} & =\sqrt{\frac{\mu_{e}\,L_{c}^{2}}{\eta_{1}}},\qquad c_{m2}=\sqrt{\frac{\mu_{e}\,L_{c}^{2}}{\eta_{2}}},\qquad c_{m3}=\sqrt{\frac{\mu_{e}\,L_{c}^{2}}{\eta_{3}}}\qquad\ \ c_{p}=\sqrt{\frac{\lambda_{e}+2\mu_{e}}{\rho}},\qquad c_{s}=\sqrt{\frac{\mu_{e}+\mu_{c}}{\rho}}.\nonumber 
\end{align}

The dispersion curves for the weighted relaxed micromorphic model
with free and gradient micro-inertia can hence be obtained as the
solutions $\omega=\omega(k)$ of the algebraic equations 
\begin{equation}
\underbrace{\mathrm{det}\,A_{1}(\omega,k)=0,}_{\text{longitudinal}}\qquad\qquad\underbrace{\mathrm{det}\,A_{2}(\omega,k)=\mathrm{det}\,A_{3}(\omega,k)=0,}_{\text{transverse}}\qquad\qquad\underbrace{\mathrm{det}\,A_{4}(\omega,k)=0.}_{\text{uncoupled}}\label{Dispersion}
\end{equation}

\begin{figure}[H]
\begin{centering}
\includegraphics[height=6.2cm]{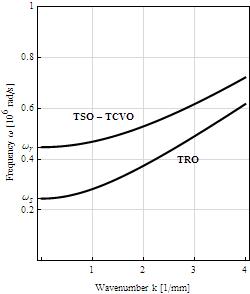} \includegraphics[height=6.2cm]{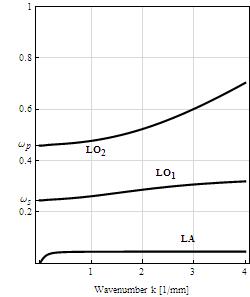}
\includegraphics[height=6.2cm]{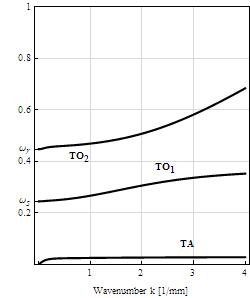} 
\par\end{centering}
\caption{\label{RelClassIn}Characteristic dispersion curves of the weighted
relaxed micromorphic model obtained with the tentative values of the
parameters shown in Tab. \ref{ParametersValues}.}
\end{figure}

We show in Fig. \ref{RelClassIn} the characteristic dispersion curves
that can be obtained via the weighted relaxed micromorphic model.
In such figures the following acronymshave been used accordingly to
the preceding papers on this subject: 
\begin{itemize}
\item TRO: transverse rotational optic, 
\item TSO: transverse shear optic, 
\item TCVO: transverse constant-volume optic, 
\item LA: longitudinal acoustic, 
\item LO$_{1}$-LO$_{2}$: $1^{st}$ and $2^{nd}$ longitudinal optic, 
\item TA: transverse acoustic, 
\item TO$_{1}$-TO$_{2}$: $1^{st}$ and $2^{nd}$ transverse optic. 
\end{itemize}
To draw the curves in Fig. \ref{RelClassIn}, we chose the tentative
values of the characteristic parameters shown in table \ref{ParametersValues}.
From the observation of such curves many features which are important
for the subsequent fitting of the parameters can be identified:
\begin{itemize}
\item The first main characteristic of the relaxed micromorphic model is
that the longitudinal and transverse acoustic waves have an horizontal
asymptote which has been seen to be essential for the description
of band-gaps in enriched continua \cite{dagostino2016panorama,madeo2014band,madeo2015wave,madeo2016complete,madeo2016reflection}.
Moreover, it has been shown in \cite{dagostino2016panorama} that
the slopes (close to the origin) of the longitudinal and transverse
acoustic curves can be expressed in terms of the parameters of the
relaxed model as 
\begin{gather}
\sqrt{\frac{3\,\lambda_{e}\,\lambda_{\mathrm{micro}}\,(\mu_{e}+\mu_{\mathrm{micro}})+2\mu_{e}\,\mu_{\mathrm{micro}}\left[3(\lambda_{e}+\lambda_{\mathrm{micro}})+2(\mu_{e}+\mu_{\mathrm{micro}})\right]+2\lambda_{\mathrm{micro}}\,\mu_{e}^{2}+2\lambda_{e}\,\mu_{\mathrm{micro}}^{2}}{\rho\,(\mu_{e}+\mu_{\mathrm{micro}})\left(3(\lambda_{e}+\lambda_{\mathrm{micro}})+2(\mu_{e}+\mu_{\mathrm{micro}})\right)}},\nonumber \\
\label{eq:SlopesLT}\\
\sqrt{\frac{\mu_{e}\,\mu_{\mathrm{micro}}}{\rho\,(\mu_{e}+\mu_{\mathrm{micro}})}}.\nonumber 
\end{gather}
The slope of the longitudinal and transverse acoustic curves will
be seen to be a primordial feature for the fitting of the relaxed
micromorphic model on real metamaterials. Moreover, such slopes provide
a way to compare the relaxed micromorphic model to classical Cauchy
elasticity for low frequencies (high wavelengths), since they are
a measure of the apparent macroscopic stiffnesses of the considered
metamaterial. As a matter of fact, it has been shown in \cite{barbagallo2016transparent}
that, using simple homogenization arguments, if one sets
\begin{equation}
\mu_{\mathrm{macro}}=\frac{\mu_{e}\mu_{\mathrm{micro}}}{\mu_{e}+\mu_{\mathrm{micro}}},\qquad\lambda_{\mathrm{macro}}=\frac{1}{3}\frac{\left(2\mu_{e}+3\lambda_{e}\right)\left(2\mu_{\mathrm{micro}}+3\lambda_{\mathrm{micro}}\right)}{2\left(\mu_{e}+\mu_{\mathrm{micro}}\right)+3\left(\lambda_{e}+\lambda_{\mathrm{micro}}\right)}-\frac{2}{3}\frac{\mu_{e}\mu_{\mathrm{micro}}}{\mu_{e}+\mu_{\mathrm{micro}}},\label{eq:Micro_Macro}
\end{equation}
then the slopes of the longitudinal and transverse acoustic curves
can be rewritten as
\[
\sqrt{\frac{\lambda_{\mathrm{macro}}+2\mu_{\mathrm{macro}}}{\rho}},\qquad\sqrt{\frac{\mu_{\mathrm{macro}}}{\rho}}\,.
\]
Since $\lambda_{\mathrm{macro}}$ and $\mu_{\mathrm{macro}}$ represent
the apparent macroscopic parameters of the metamaterial that we want
to describe via our enriched model, such identification of the slopes
of the acoustic curves in terms of the macroscopic elastic parameters
allows a direct comparison with classical elasticity when considering
small frequencies.
\item The second main feature of the dispersion curves that we can point
out is the presence of the cut-off frequencies $\omega_{s}$, $\omega_{r}$
and $\omega_{p}$ that are defined in \eqref{eq:characteristic_quantities}
as functions of the parameters of the weighted relaxed micromorphic
model. From their definition, we can immediately remark that each
of these cut-offs depends on a different free micro-inertia parameter,
so that each cut-off, can be identified to be related, at low wavenumbers,
to a specific vibration mode. More precisely, $\omega_{s}$ being
related to $\eta_{1}$, which in turns can be seen to be the micro-inertia
associated to the $\dev\sym$ part of $\p_{,t}$, is the cut-off of
a vibration mode initially (for low wavenumbers) associated to micro-distorsions.
Analogously, $\omega_{r}$ is related to $\eta_{2}$, which is the
micro-inertia associated to the $\skew$ part of $\p_{,t}$ and can
then be interpreted to be the cut-off of a mode initially associated
to micro-rotations. Finally $\omega_{p}$ is related to $\eta_{3}$
which is the micro-inertia associated to the $\tr$ part of $\p_{,t}$
so that it can be interpreted to be the cut-off of a mode initially
associated to volume variations. For higher wavenumbers the vibration
modes associated to each dispersion curve can vary and coupled micro-vibrational
modes can be observed on each branch. We can here underline the fact
that the splitting of the tensor $P_{,t}$ by means of the use of
its Cartan-Lie decomposition, allowing the introduction of three independent
micro-inertia parameters $\eta_{1}$, $\eta_{2}$ and $\eta_{3}$,
is fundamental to have the possibility of a reasonable fitting of
the dispersion curves on real metamaterials. Indeed, the fact of having
the freedom of moving independently the three cut-offs by varying
the value of the parameters $\eta_{1}$, $\eta_{2}$ and $\eta_{3}$
will be seen to be a fundamental feature for the fitting on real metamaterials
that we propose afterwards. 
\item The third characteristic of the weighted relaxed micromorphic model
is that related to the presence of a gradient micro-inertia. The effect
of the parameters $\bar{\eta}_{1}$, $\bar{\eta}_{2}$ and $\bar{\eta}_{3}$
on the dispersion curves is that of flattening some of the longitudinal
and transverse optic curves which can eventually take horizontal asymptotes.
As it has been shown in \cite{madeo2016role}, the gradient micro-inertia
parameters have no role on the uncoupled curves, while they provide
the aforementioned flattening effect for the longitudinal and transverse
waves. In particular, the parameter $\bar{\eta}_{1}$ has been seen
to have no specific effect on the dispersion curves, while the parameters
$\bar{\eta}_{3}$ and $\bar{\eta}_{2}$ have been seen to be separately
responsible for the flattening of the longitudinal and transverse
waves, respectively.
\item We finally point out that if considering a 2D problem, the uncoupled
waves have not to be taken into account. Indeed, with reference to
Eqs. (\ref{Shear}-\ref{VolumeVariation}), the uncoupled equations
govern the evolution of the quantities of the quantities $P_{(23)}$,
$P_{[23]}$ and $P^{V}$. Under the hypothesis of plane micro-strain,
the variables $P_{(23)}$ and $P_{[23]}$ are automatically vanishing,
while the variable $P^{V}$ reduces, according to its definition,
to the variable $P_{22}$ which results to be known when the longitudinal
quantities $P^{D}$ and $P^{S}$ are determined. For these reasons,
the uncoupled waves will not appear in the fitting procedure on the
2D metamaterials considered in the following sections.
\end{itemize}
\begin{table}[H]
\begin{centering}
\begin{tabular}{|c|c|c|}
\hline 
Parameter  & Value  & Unit\tabularnewline
\hline 
\hline 
$\me$  & $200$  & $\mathrm{MPa}$\tabularnewline
\hline 
$\lle=2\me$  & 400  & $\mathrm{MPa}$\tabularnewline
\hline 
$\mc=5\me$  & $ $1000  & $\mathrm{MPa}$\tabularnewline
\hline 
$\mh$  & 100  & $\mathrm{MPa}$\tabularnewline
\hline 
$\lh$  & $100$  & $\mathrm{MPa}$\tabularnewline
\hline 
$L_{c}\ $  & $1$  & $\mathrm{mm}$\tabularnewline
\hline 
$\rho$  & $2000$  & $\mathrm{kg/m^{3}}$\tabularnewline
\hline 
$\eta$  & $ $$10^{-2}$  & $\mathrm{kg/m^{3}}$\tabularnewline
\hline 
$\overline{\eta}$  & $ $$10^{-1}$  & $\mathrm{kg/m^{3}}$\tabularnewline
\hline 
\end{tabular}\quad{}\quad{}\quad{}\quad{}%
\begin{tabular}{|c|c|c|}
\hline 
Parameter  & Value  & Unit\tabularnewline
\hline 
\hline 
$\lm$  & $82.5$  & $\mathrm{MPa}$\tabularnewline
\hline 
$\mm$  & $66.7$  & $\mathrm{MPa}$\tabularnewline
\hline 
$E_{\mathrm{macro}}$  & $170$  & $\mathrm{MPa}$\tabularnewline
\hline 
$\nu_{\mathrm{macro}}$  & $0.28$  & $-$\tabularnewline
\hline 
\end{tabular}
\par\end{centering}
\caption{\label{ParametersValues}Tentative values of the parameters used in
the numerical simulations (left) and corresponding values of the Lamï¿œ
parameters and of the Young modulus and Poisson ratio (right), for
the formulas needed to calculate the homogenized macroscopic parameters
starting from the microscopic ones, see \cite{barbagallo2016transparent}.}
\end{table}

In the following section, we will target a precise microstructure
which is known to show band-gaps and we will use the weighted relaxed
micromorphic model presented before to obtain, by inverse approach,
the values of the elastic parameters of the model for this particular
microstructure.

\section{Fitting the parameters of the weighted relaxed micromorphic model
on real band-gap metamaterials}

\begin{figure}[H]
\begin{centering}
\begin{tabular}{cc}
\includegraphics[scale=0.7]{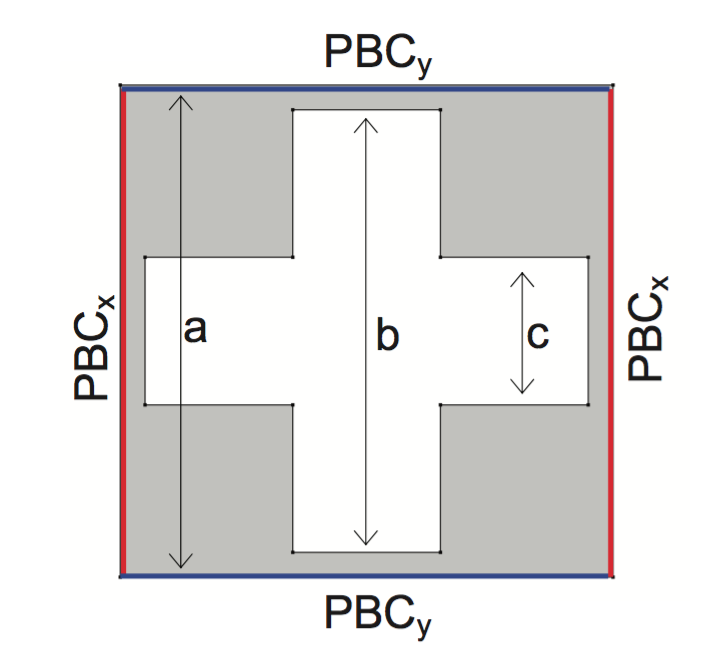} & \includegraphics[scale=0.5]{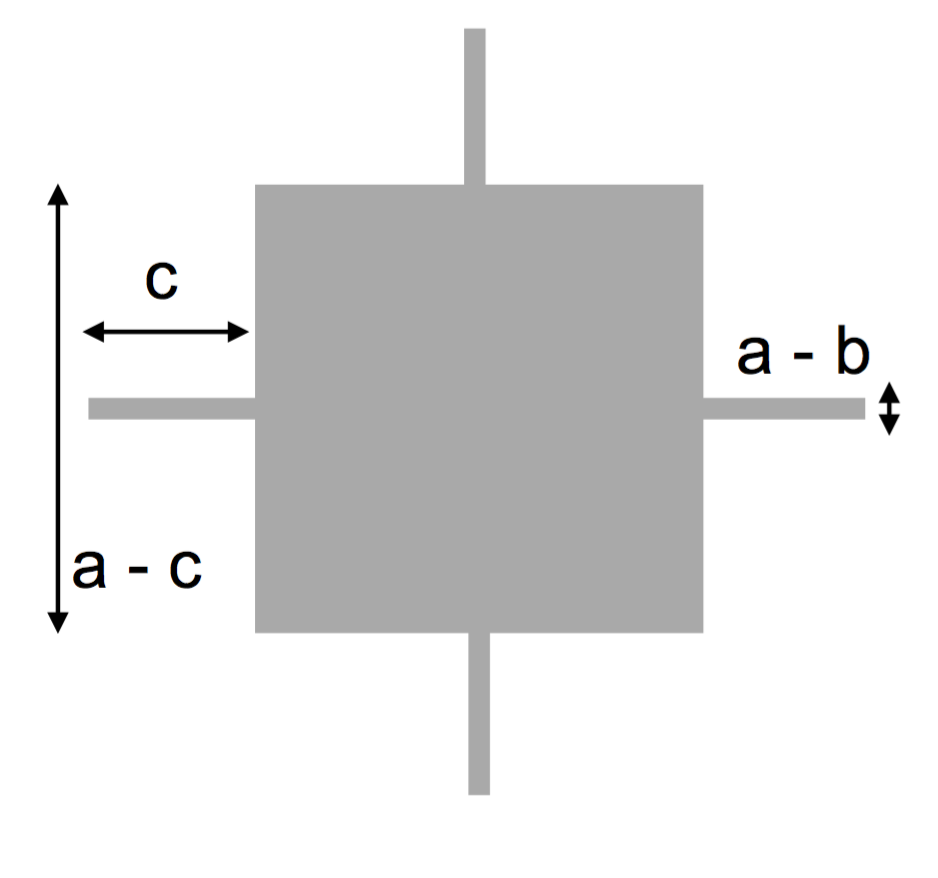}\tabularnewline
(a) & (b)\tabularnewline
\end{tabular}
\par\end{centering}
\caption{\label{fig:micro_1} (a): Topology and elastic properties of a specific
metamaterial exhibiting band-gaps. The grey region is filled by aluminum,
the white region is empty. (b): Equivalent periodic building block
of the considered phononic structure allowing a more direct identification
of the mass-spring mechanism at the microstructural level.}

\end{figure}

The object of this section is that of starting to fit some of the
parameters of the relaxed micromorphic model on real band-gap metamaterials.
For this reason, we start considering the two-dimensional metamaterial
shown in Fig. \ref{fig:micro_1} which is known to exhibit band-gap
behaviors with respect to elastic wave propagation.

\begin{table}[H]
\begin{centering}
\begin{tabular}{|c|c|c|c|c|c|}
\hline 
a & b & c & $\rho$ & $E$ & $\nu$\tabularnewline
\hline 
{[}$\mathrm{mm}${]} & {[}$\mathrm{mm}${]} & {[}$\mathrm{mm}${]} & {[}$\mathrm{kg/m^{3}}${]} & {[}$\mathrm{GPa}${]} & {[}$-${]}\tabularnewline
\hline 
\hline 
$1$ & $0.9$ & $0.3$ & $2700$ & $70$ & $0.33$\tabularnewline
\hline 
\end{tabular}
\par\end{centering}
\caption{Values of the elastic parameters of the base material (aluminum) and
geometric parameters relative to the unit cell shown in Fig. \ref{fig:micro_1}.}

\end{table}

The unit cell shown in Fig. \ref{fig:micro_1} (a) is clearly equivalent
to the one presented in Fig. \ref{fig:micro_1} (b) which can be more
directly interpreted as a mass-spring system. This mass-spring representation
can be useful when one wants to apply classical homogenization methods
to the considered periodic structure. Indeed, following upscaling
methods like the ones presented in \cite{aurialt2012long,boutin2003homogenisation,boutin2010generalized},
some informations relative to the micro-mechanisms activated at low
frequency (large wavelengths allowing the so-called separation of
scale hypothesis) could be disclosed and this would permit a clearer
interpretation of some of the parameters of the homogenized relaxed
micromorphic model. The problem of establishing such first micro-macro
relationships will be the aim of a subsequent work, since we are interested
here in the description of the metamaterial at the macroscopic level
without restrictions on the frequency (or, equivalently on the wavelength).
Indeed, with the methods presented here, we are able to recover the
behavior of the metamaterial when excited at wavelengths which go
down to the size of the periodic unit cell.

\begin{figure}
\begin{centering}
\begin{tabular}{c}
\quad{}\quad{}\quad{}\quad{}\includegraphics[scale=0.8]{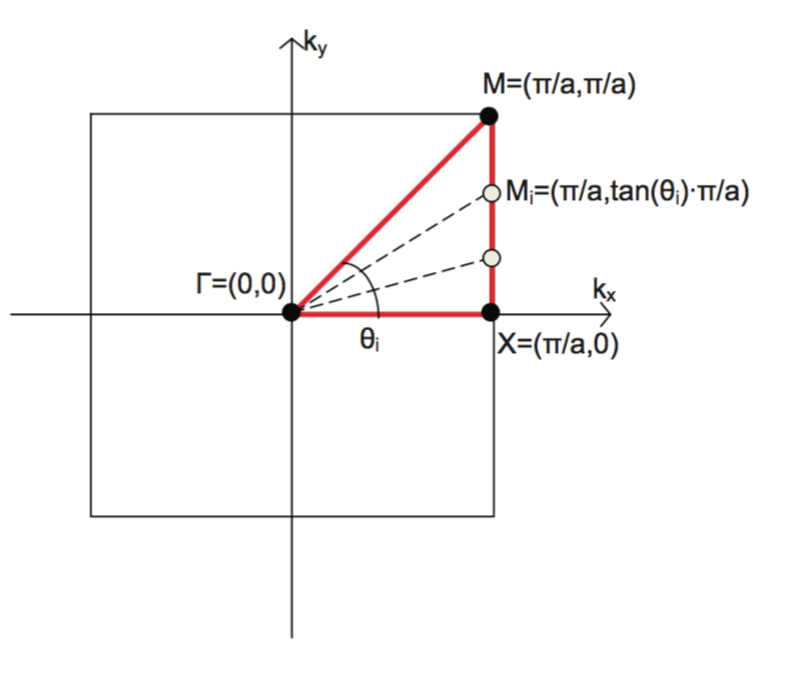}\tabularnewline
\end{tabular}
\par\end{centering}
\caption{\label{fig:brillouin}Possible wave paths spanning in the Brillouin
zone. We indicate by $\theta_{i}$ the angle that the direction of
propagation of the wave $i$ forms with the $x$ direction.}
\end{figure}

\begin{figure}[H]
\begin{centering}
\includegraphics[scale=0.45]{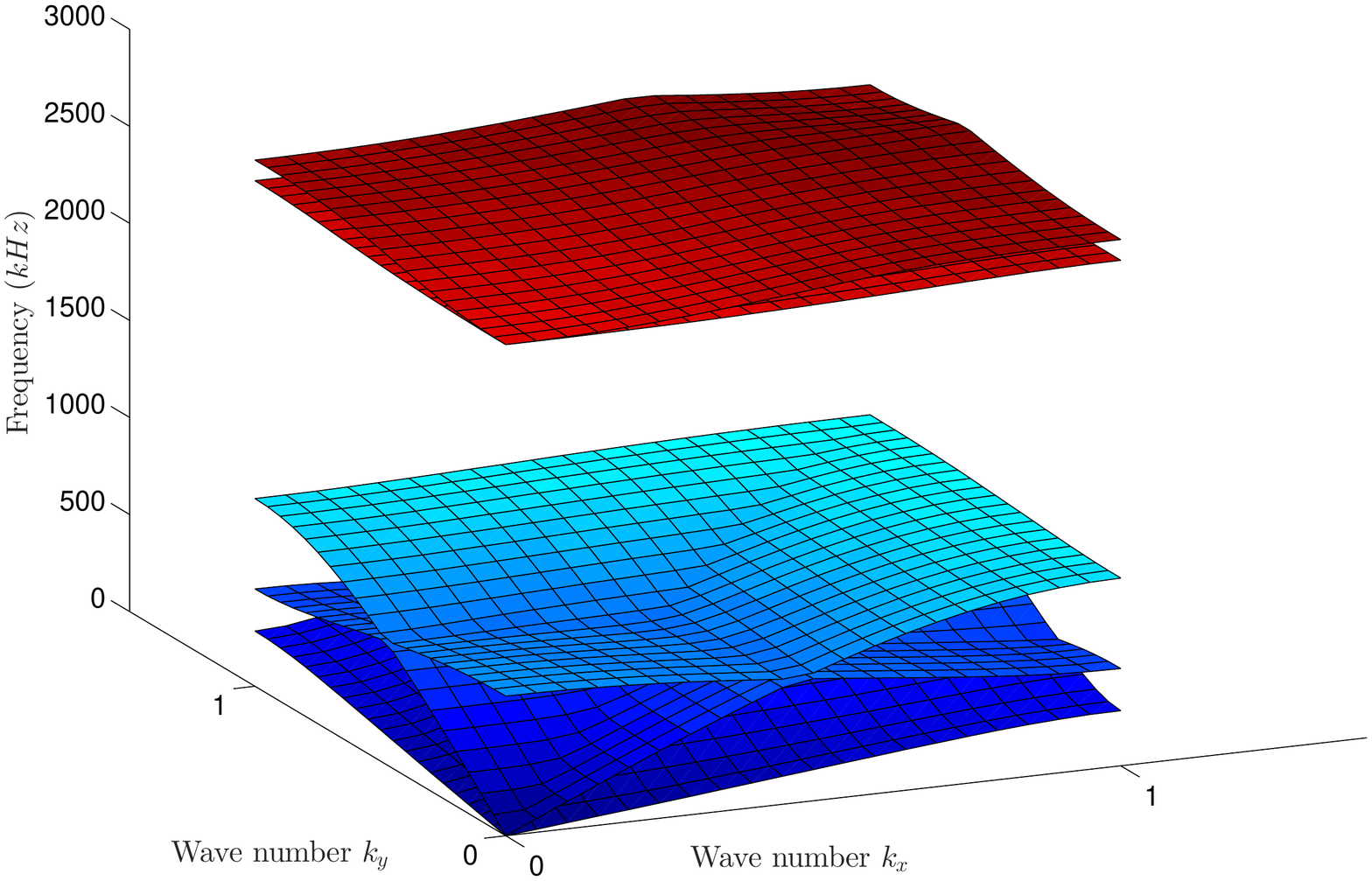}
\par\end{centering}
\caption{\label{fig:Dispersion-surfaces}Dispersion surfaces as obtained by
the application of Bloch boundary conditions to the unit cell shown
in Fig. \ref{fig:micro_1}.}

\end{figure}

A Bloch analysis has been performed (see e.g. \cite{collet2011floquet,fan2016energy,fan2016wave})
on such metamaterial for waves traveling in the $\Gamma X$, $\Gamma M$
and $XM$ of the usual Brillouin zone, as well as for waves traveling
in arbitrary directions within the Brillouin zone (see Fig. \ref{fig:brillouin}).
As a result of such analysis, the dispersion surfaces for the considered
metamaterial have been obtained and are shown in Fig. \ref{fig:Dispersion-surfaces}.
It can be noticed from Fig. \ref{fig:Dispersion-surfaces} that considering
an arbitrary wave-vector of components $(k_{x},k_{y})$ (which means
a wave traveling in an arbitrary direction with associated angle $\theta_{i}$),
very few changes can be observed on the dispersion curves, especially
for those which are bounding the band-gap region. This means that
the behavior of the material can be considered not far to be isotropic,
even if some low degree of anisotropy of course exists. In this paper,
we hence suppose that the considered metamaterial has an isotropic
behavior, so that it is reasonable to use the constitutive form \eqref{eq:Ener-General}
of the energy to model it through our enriched continuum model.

\begin{figure}[H]
\begin{centering}
\includegraphics[scale=0.5]{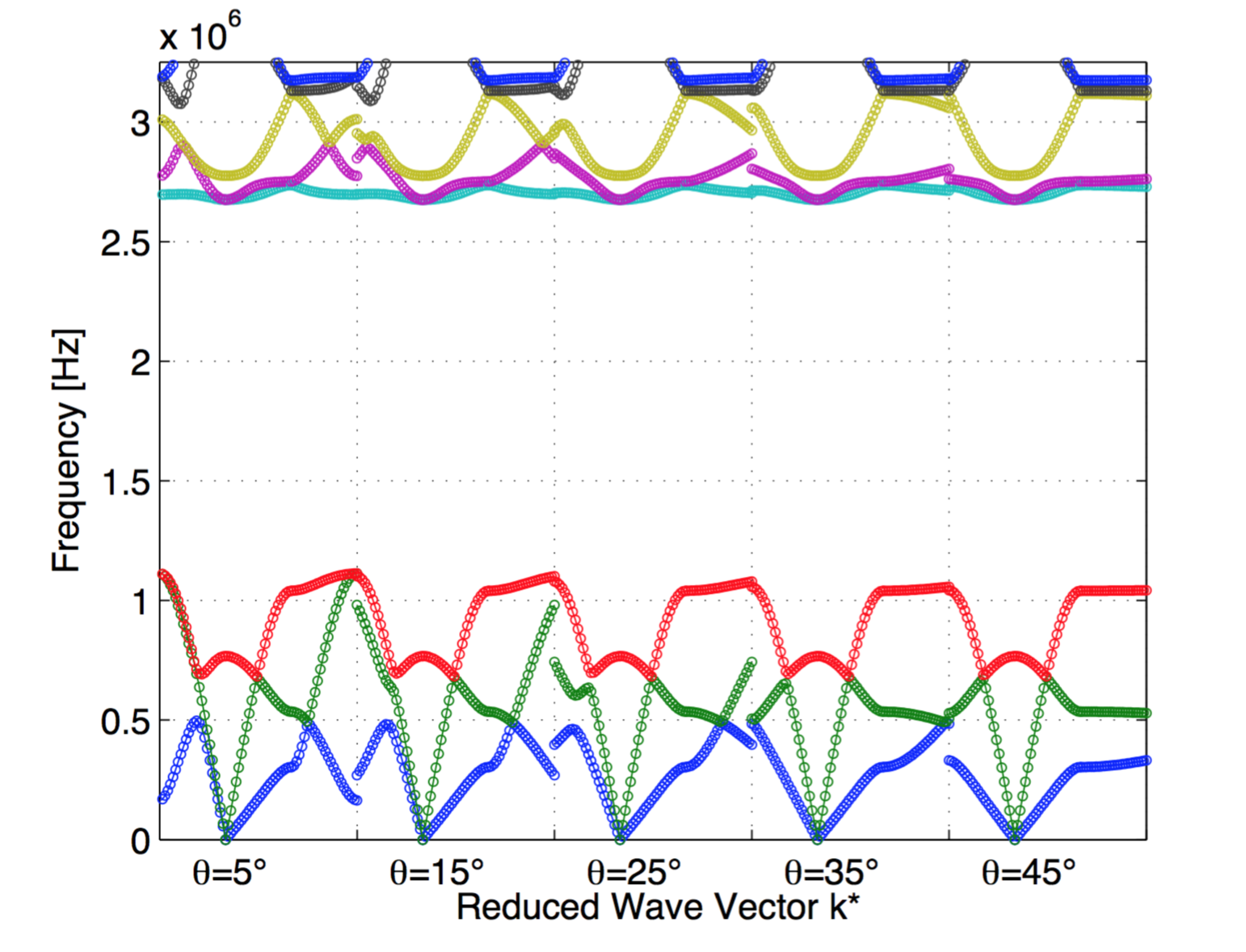}
\par\end{centering}
\caption{\label{fig:Dispersion-curves}Dispersion curves relative to waves
propagating in different directions $\theta_{i}$ within the considered
metamaterial. An almost isotropic behavior can be detected, especially
concerning the band-gap region.}
\end{figure}

To better illustrate our choice of using an isotropic model for the
considered metamaterial, we present in Fig. \ref{fig:Dispersion-curves}
the dispersion curves for wave vectors spanning within the Brillouin
zone and having different traveling directions. 
 It can be indeed inferred from Fig. \ref{fig:Dispersion-curves}
that if some small changes can be observed in the acoustic waves when
changing the direction of propagation, almost no change intervenes
for the optic curves, in particular for those which bound the band-gap
(red and light blue). This observation strengthen our choice of using
an isotropic relaxed micromorphic model to characterize the mechanical
behavior of the considered metamaterial. Even if the error introduced
here using an isotropic model for the considered metamaterial is reasonably
small, the problem of studying wave propagation in the anisotropic
framework is worth for increasing the number of metamaterials that
can be effectively modeled via the relaxed micromorphic model. The
case of wave propagation in relaxed micromorphic continua in the anisotropic
setting will be treated in further works.

\begin{figure}[H]
\begin{centering}
\begin{tabular}{cc}
\negthickspace{}\negthickspace{}\negthickspace{}\negthickspace{}\negthickspace{}\negthickspace{}\negthickspace{}\negthickspace{}\negthickspace{}\negthickspace{}\negthickspace{}\negthickspace{}\negthickspace{}\negthickspace{}\negthickspace{}\negthickspace{}\negthickspace{}\includegraphics[scale=0.35]{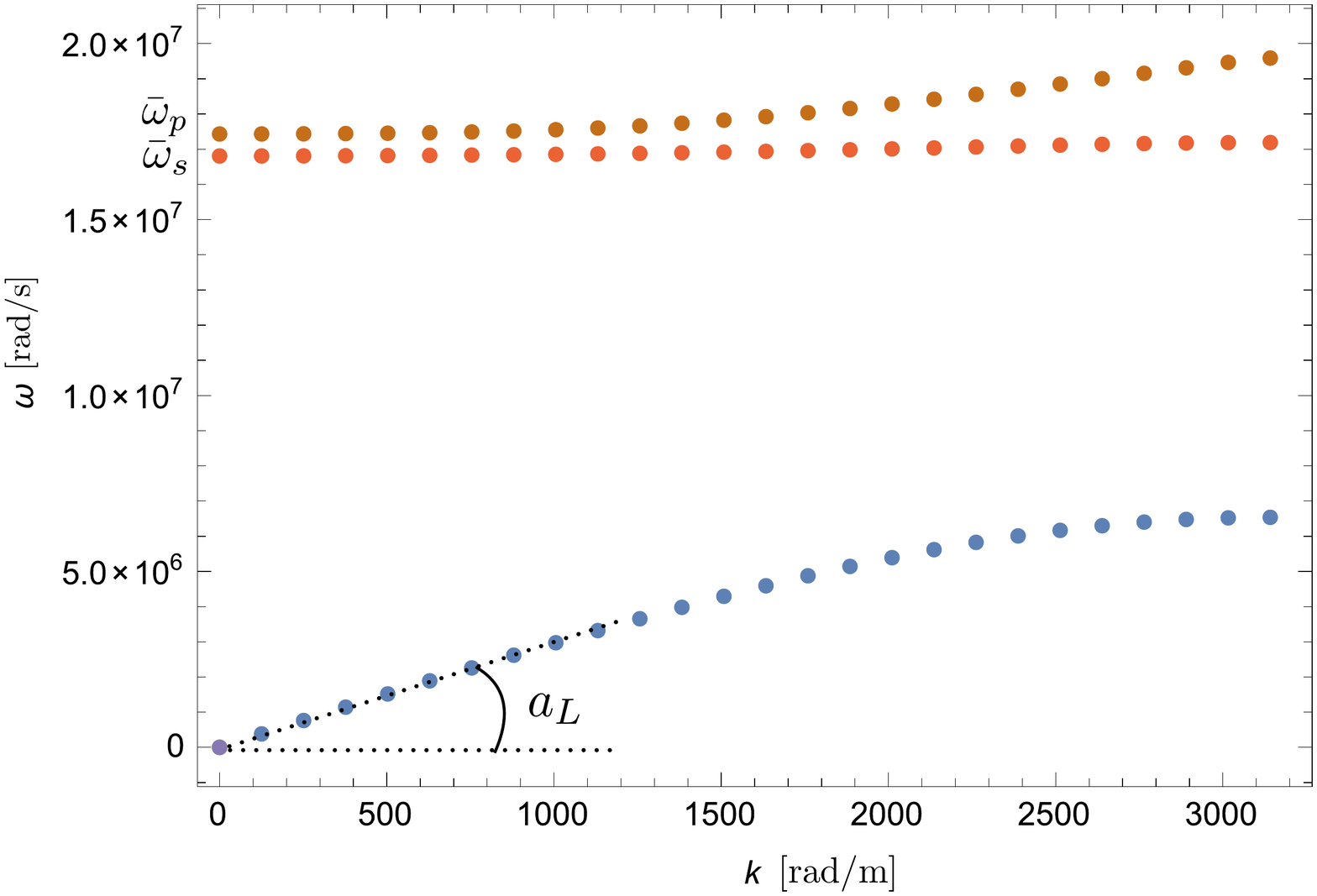} & \negthickspace{}\negthickspace{}\negthickspace{}\negthickspace{}\negthickspace{}\negthickspace{}\negthickspace{}\negthickspace{}\negthickspace{}\negthickspace{}\negthickspace{}\negthickspace{}\negthickspace{}\negthickspace{}\negthickspace{}\negthickspace{}\negthickspace{}\negthickspace{}\includegraphics[scale=0.35]{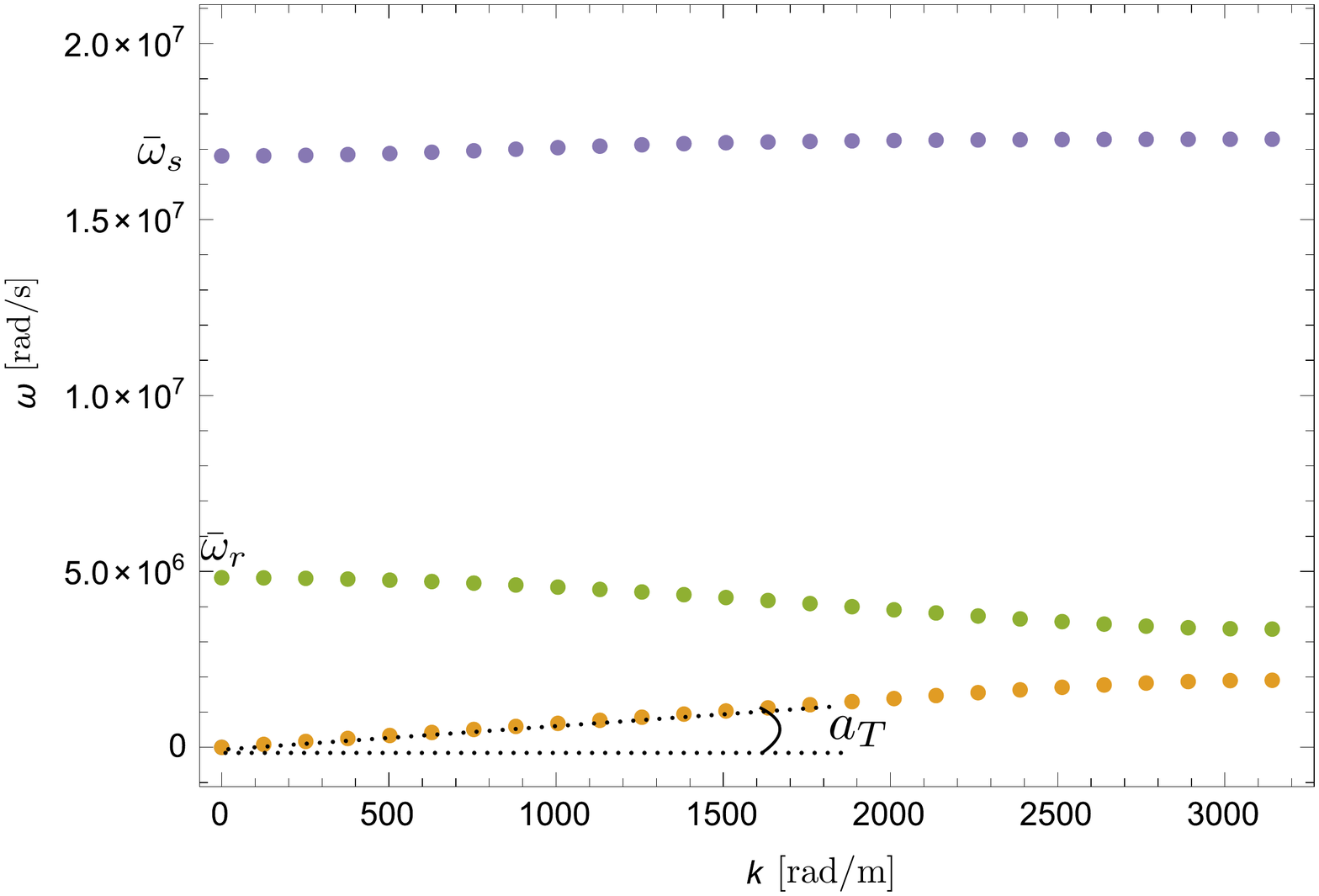}\tabularnewline
\end{tabular}
\par\end{centering}
\centering{}\caption{\label{fig:Dispersion_for_fitting}Dispersion curves for the path
$\Gamma X$ ($\theta=0$) obtained by means of the code COMSOL$^{\circledR}$
applying Bloch boundary conditions. Three cut-off frequencies $\bar{\omega}_{s}$,
$\bar{\omega}_{p}$ and $\bar{\omega}_{r}$ can be identified.}
\end{figure}

In Figure \ref{fig:Dispersion_for_fitting}, we show the dispersion
curves for the considered metamaterial relative to a wave traveling
in the $\Gamma X$ direction. Such dispersion curves will be those
used for the fitting of the parameters of the weighted relaxed micromorphic
model. In virtue of the remarks concerning the low anisotropy of the
considered metamaterial, the estimate of the parameters on the basis
of such curves will be sufficient to start characterizing the metamaterial
itself with a reasonable precision. As it was previously said, nevertheless,
the fitting on the basis of the dispersion curves alone is not sufficient
for a definitive identification of all the elastic coefficients of
the considered metamaterial and more accurate fitting procedures based
on the transmission coefficient have to be introduced.

\subsection{Fitting procedure}

The goal that we want to reach finally is that of characterizing the
material behavior of the metamaterial shown in Fig.\ref{fig:micro_1}
by estimating the value of the elastic parameters appearing in the
constitutive expression \eqref{eq:Ener-General} of the strain energy
density. To this aim, we present here the fitting procedure that has
been used to superimpose the dispersion curves obtained via the relaxed
micromorphic model as solutions of equations \eqref{Dispersion} to
the dispersion curves obtained via the Bloch analysis performed by
means of the software COMSOL$^{\circledR}$.
\begin{itemize}
\item The first step has been that of fixing the value of the apparent density
$\rho$ of the unit cell. To this purpose, we started calculating
the mass $M_{\mathrm{al}}$ of aluminum which is present in each unit
cell. With reference to Fig. \ref{fig:micro_1}, it is easy to notice
that the volume $V_{\mathrm{al}}$ of aluminum present in the unit
cell is estimated to be $V_{\mathrm{al}}=4.9\times10^{-7}\:\mathrm{m}^{3}$.
Since the density $\rho_{\mathrm{micro}}$ of the aluminum is known,
the mass of aluminum inside the unit cell is easily calculated as
$M_{\mathrm{al}}=\rho_{\mathrm{micro}}\,V_{\mathrm{al}}=0.001323\:\mathrm{Kg}.$
Being the volume $V_{\mathrm{macro}}$ of the unit cell known, the
apparent density of the unit cell is estimated to be $\rho=M_{\mathrm{al}}/V_{\mathrm{macro}}=1323\:\mathrm{kg/m^{3}}$.
\end{itemize}
Once the apparent density of the metamaterial has been fixed, the
5 elastic coefficients $\lambda_{e}$, $\mu_{e}$, $\lambda_{\mathrm{micro}}$,
$\mu_{\mathrm{micro}}$, $\mu_{c}$ plus the 6 micro-inertia parameters
$\eta_{i}$, $\bar{\eta}_{i}$, $i=\left\{ 1,2,3\right\} $ and characteristic
length $L_{c}$ still need to be determined.
\begin{itemize}
\item We started setting the characteristic length $L_{c}$ to be vanishing
in order to fit the remaining elastic coefficients. Indeed, it was
already shown in \cite{madeo2016first} that $L_{c}$ is related to
the non-locality of the metamaterials and its determination is very
delicate since, energetically, a non-vanishing $L_{c}$ brings only
small corrections to the case with $L_{c}=0$. The value of the other
parameters is thus not affected if they are fitted by setting, in
a first approximation, $L_{c}=0$. We anticipate the fact that, in
the present paper, a reliable value of the non-locality cannot be
determined, since the fitting of the parameters on the dispersion
curves alone is a tool which is not accurate enough to accomplish
this task. In other words, a fitting based on the dispersion curves,
if valuable to estimate the elastic parameters, is not sufficient
to calibrate the characteristic length which measures non-localities.
To this aim, more precise fitting procedures have to be used, such
as that of fitting the transmission coefficient obtained via the relaxed
micromorphic model to that which is observed experimentally or which
is issued by a numerical simulation which accounts for all the elements
of the microstructure. In all the remainder of this paper we thus
retain the value $L_{c}=0$, simultaneously pointing out that extra
investigations of the type presented in \cite{madeo2016first} are
needed to determine the precise value of such parameter.
\end{itemize}
In order to fit the remaining parameters, we started imposing some
conditions that such parameters must necessarily satisfy with the
aim of reducing the number of free parameters of the model. In order
to obtain such conditions, we remark that:
\begin{itemize}
\item With reference to the dispersion curves of the relaxed micromorphic
model shown in Fig. \ref{RelClassIn}, the three cut-off frequencies
$\omega_{s}$, $\omega_{r}$ and $\omega_{p}$ can be calibrated to
coincide with those issued by means of the Bloch analysis shown in
Fig. \ref{fig:Dispersion_for_fitting}. In this latter, we can notice
that the lower cut-off frequency $\bar{\omega}_{r}$ is associated
to rotational modes and is thus identified with $\omega_{r}$, while
the higher cut-off $\bar{\omega}_{p}$ is identified with $\omega_{p}$.
Two waves (one longitudinal and one transverse) are seen to have approximately
the same cut-off $\bar{\omega}_{s}$ which is thus identified with
the characteristic frequency $\text{\ensuremath{\omega}}_{s}$ of
the relaxed micromorphic model. Using the definitions \eqref{eq:characteristic_quantities}
we can hence introduce the following equalities that must be satisfied
by the parameters of the relaxed micromorphic model (we recall that
$\bar{\omega}_{r}$, $\bar{\omega}_{s}$ and $\bar{\omega}_{p}$ are
known) 
\begin{equation}
\sqrt{\frac{(3\,\lambda_{e}+2\,\mu_{e})+(3\,\lh+2\,\mh)}{\eta_{3}}}=\bar{\omega}_{p},\qquad\sqrt{\frac{2\,(\mu_{e}+\mh)}{\eta_{1}}}=\bar{\omega}_{s},\qquad\sqrt{\frac{2\,\mu_{c}}{\eta_{2}}}=\bar{\omega}_{r}.\label{eq:Imposizione_1}
\end{equation}
\item We want also to impose that the slope of the acoustic curves of the
micromorphic model is the same as that obtained via the Bloch analysis.
To do so, with reference to Eqs. \eqref{eq:SlopesLT} we impose that
\begin{gather}
\sqrt{\frac{3\,\lambda_{e}\,\lambda_{\mathrm{micro}}\,(\mu_{e}+\mu_{\mathrm{micro}})+2\mu_{e}\,\mu_{\mathrm{micro}}\left[3(\lambda_{e}+\lambda_{\mathrm{micro}})+2(\mu_{e}+\mu_{\mathrm{micro}})\right]+2\lambda_{\mathrm{micro}}\,\mu_{e}^{2}+2\lambda_{e}\,\mu_{\mathrm{micro}}^{2}}{\rho\,(\mu_{e}+\mu_{\mathrm{micro}})\left(3(\lambda_{e}+\lambda_{\mathrm{micro}})+2(\mu_{e}+\mu_{\mathrm{micro}})\right)}}=a_{L},\nonumber \\
\label{eq:Imposizione_2}\\
\sqrt{\frac{\mu_{e}\,\mu_{\mathrm{micro}}}{\rho\,(\mu_{e}+\mu_{\mathrm{micro}})}}=a_{T},\nonumber 
\end{gather}
where $a_{L}$ and $a_{T}$ are the numerical values of the slopes
of the acoustic longitudinal and transverse branches, respectively,
as obtained via the Bloch analysis.
\end{itemize}
By solving equations \eqref{eq:Imposizione_1}, \eqref{eq:Imposizione_2}
with respect to a suitable subset of the unknown parameters, we find
the following solution
\begin{gather}
\mu_{c}=\frac{\bar{\omega}_{r}^{2}}{2}\eta_{2},\qquad\mu_{e}=\frac{\rho\,a_{T}^{2}\,\mu_{\mathrm{micro}}}{\mu_{\mathrm{micro}}-\rho a_{T}^{2}},\qquad\lambda_{e}=-\lambda_{\mathrm{micro}}+\frac{2}{3}\frac{\mu_{\mathrm{micro}}^{2}}{-\mu_{\mathrm{micro}}+\rho\,a_{T}^{2}}+\frac{\left(3\,\lambda_{\mathrm{micro}}+2\,\mu_{\mathrm{micro}}\right)^{2}}{3\,\left(3\,\lambda_{\mathrm{micro}}+2\,\mu_{\mathrm{micro}}-3\rho\,a_{L}^{2}+4\rho a_{T}^{2}\right)},\nonumber \\
\label{eq:partial_Solution}\\
\eta_{1}=\frac{2\mu_{\mathrm{micro}}^{2}}{\left(\mu_{\mathrm{micro}}-\rho a_{T}^{2}\right)\bar{\omega}_{s}^{2}},\qquad\eta_{3}=\frac{\left(3\lambda_{\mathrm{micro}}+2\mu_{\mathrm{micro}}\right)^{2}}{\left(3\lambda_{\mathrm{micro}}+2\mu_{\mathrm{micro}}-3\rho\,a_{L}^{2}+4\rho\,a_{T}^{2}\right)\bar{\omega}_{p}^{2}}.\nonumber 
\end{gather}
The parameters of the model which remain still free are thus $\lambda_{\mathrm{micro}}$,
$\mu_{\mathrm{micro}}$, $\eta_{2}$, $\bar{\eta}_{1}$, $\bar{\eta}_{2}$
and $\bar{\eta}_{3}$. The last two equations can also be inverted
in order to find $\lambda_{\mathrm{micro}}$ and $\mu_{\mathrm{micro}}$
as functions of $\eta_{1}$ and $\eta_{3}$, which is a more desirable
situation to perform the fitting procedure. In this way, the free
parameters to be varied for the fitting procedures would be only the
free and gradient micro inertias $\eta_{1}$, $\eta_{2}$, $\eta_{3}$,
$\bar{\eta}_{1}$, $\bar{\eta}_{2}$ and $\bar{\eta}_{3}$. Indeed,
regarding the last two equations \eqref{eq:partial_Solution} in such
a way to solve them in terms of $\mu_{\mathrm{micro}}$ and $\lambda{}_{\mathrm{micro}}\,$,
we get the following solutions 
\begin{align}
\mu_{\mathrm{micro}} & =\frac{1}{4}\left(\eta_{1}\bar{\omega}_{s}^{2}\pm\sqrt{\eta_{1}\bar{\omega}_{s}^{2}\,(-8\,\rho\,a_{T}^{2}+\eta_{1}\bar{\omega}_{s}^{2})}\right),\nonumber \\
\label{eq:conditions}\\
\lambda_{\mathrm{micro}} & =\frac{1}{6}\left(-4\mu_{\mathrm{micro}}+\eta_{3}\bar{\omega}_{p}^{2}\pm\sqrt{\eta_{3}\bar{\omega}_{p}^{2}\,(-12\,\rho\,a_{L}^{2}+16\,\rho\,a_{T}^{2}+\eta_{3}\bar{\omega}_{p}^{2})}\right).\nonumber 
\end{align}
This means that, for any arbitrary value of $\eta_{1}$ and $\eta_{3}$,
there are two possible values of $\mu_{\mathrm{micro}}$ and hence
four possible values of $\lambda{}_{\mathrm{micro}}\,$. In other
words, If we think to vary the values of $\eta_{1}$ and $\eta_{3}$,
four different possible solutions arise and one among them must be
chosen on the basis of experimental observations. Since, as a physical
requirement, $\mu_{\mathrm{micro}}$ and $\lambda{}_{\mathrm{micro}}\,$
must be real, then some restrictions on $\eta_{1}$ and $\eta_{3}$
have to be imposed in order to avoid complex solutions for such elastic
coefficients. It can be remarked that, on the basis of Eqs. \eqref{eq:conditions},
the following conditions must be imposed on the micro-inertias $\eta_{1}$
and $\eta_{3}$ in order to guarantee that $\mu_{\mathrm{micro}}$
and $\lambda{}_{\mathrm{micro}}\,$ take only real values
\begin{equation}
\eta_{1}>\frac{8\,\rho\,a_{T}^{2}}{\bar{\omega}_{s}^{2}},\qquad\eta_{3}>\frac{4\,\rho\left(3\,a_{L}^{2}-4\,a_{T}^{2}\right)}{\bar{\omega}_{p}^{2}}\:.\label{eq:l_m_reals}
\end{equation}
Such conditions suggest the minimal numerical values for $\eta_{1}$
and $\eta_{2}$ which must be used to start the fitting procedure.
The value of the two micro-inertias $\eta_{1}$ and $\eta_{2}$ will
be in fact slowly increased starting from their minimal value in order
to fit the dispersion curves of the relaxed micromorphic model on
those obtained via the Bloch analysis\footnote{It can be checked that the expressions \eqref{eq:conditions} for
$\mu_{\mathrm{micro}}$ and $\lambda{}_{\mathrm{micro}}\,$ together
with a choice of $\eta_{1}$ and $\eta_{3}$ complying with the conditions
\eqref{eq:l_m_reals} imply that $\mu_{\mathrm{micro}}>0$ and $3\lambda_{\mathrm{micro}}+2\mu_{\mathrm{micro}}>0$.
Moreover, such conditions on $\mu_{\mathrm{micro}}$ and $\lambda{}_{\mathrm{micro}}\,$
also imply that, given equations \eqref{eq:partial_Solution}, $\mu_{e}>0$
and $3\lambda_{e}+2\mu_{e}>0$. This means that, in the end, the only
fact of using the restrictions \eqref{eq:l_m_reals} and of additionally
imposing $\eta_{2}\geq0$, imply positive definiteness of the strain
energy density.}. 

In order to fit at best the remaining free parameters $\eta_{1}$,
$\eta_{2}$, $\eta_{3}$, $\bar{\eta}_{1}$, $\bar{\eta}_{2}$, $\bar{\eta}_{3}$,
we have performed a systematic numerical check to unveil eventual
peculiar effects of each parameter on the dispersion curves. The fitting
procedure has been performed as follows
\begin{itemize}
\item The characteristic effect of each parameter on the dispersion curves
is searched by slightly increasing each of them starting from the
initial values
\[
\eta_{1}=\frac{8\,\rho\,,a_{T}^{2}}{\bar{\omega}_{s}^{2}},\quad\eta_{2}=0,\quad\eta_{3}=\frac{4\,\rho\left(3\,a_{L}^{2}-4\,a_{T}^{2}\right)}{\bar{\omega}_{p}^{2}},\quad\bar{\eta}_{1}=0,\quad\bar{\eta}_{2}=0,\quad\bar{\eta}_{3}=0.
\]
\item It is found that $\eta_{1}$ has a visible effect on the optic curves
$LO_{1}$ and $TO_{1}$ as well as a smaller effect on the acoustic
curves $LA$ and $TA$. The parameter $\eta_{1}$ is then carefully
increased in order to reach the best possible agreement with the acoustic
curves as well as with the curves $LO_{1}$ and $TO_{1}$. A rather
good agreement is found (for one of the four possible solutions) for
all the aforementioned curves for a first tentative value of $\eta_{1}$,
except for the curve $LO_{1}$ that drastically diverge from that
issued by the discrete simulation, above all for high wavenumbers.
\item The parameter $\bar{\eta}_{3}$ is seen to have a flattening effect
on the longitudinal curves. Increasing this parameter allows a better
fitting of the curve $LO_{1}$ which was still remaining to be better
adjusted from the previous step.
\item The gradient micro-inertia parameter $\bar{\eta}_{1}$ has the effect
of flattening simultaneously the longitudinal and transverse waves.
This effect is not desirable since the transverse curves are already
well-fitted at this stage. We then set $\bar{\eta}_{1}$ to be identically
vanishing.
\item The parameter $\eta_{2}$ is seen to have an effect on the curves
$TO_{2}$ and $TA$. In particular, for some values of $\eta_{2}$,
the curve $TO_{2}$ is seen to be almost flat. This is the best agreement
that we can find with the curve issued via the Bloch wave analysis,
which is instead slightly decreasing. Up to now, the introduced formulation
of the relaxed micromorphic model does not allow to obtain dispersion
branches which decrease when increasing the wavenumber, thus the best
approximation is obtained with an almost horizontal line. We will
discuss the possibility of obtaining such decreasing curves in the
framework of the relaxed micromorphic model in subsequent works. In
order to get the best average fitting of the curve $TO_{2}$, we slightly
relax the third of conditions \eqref{eq:Imposizione_1} in order to
impose a cut-off which takes a lower value than the exact value $\bar{\omega}_{r}$.
We also remark that, indeed, the desired horizontal curve $TO_{2}$
can be obtained for an infinite set of values of $\eta_{2}$. As a
consequence, there exist an infinity of calculated values of the Cosserat
couple modulus $\mu_{c}$ (via the third of conditions \eqref{eq:Imposizione_1})
which correspond to such set of values of $\eta_{2}$. This indeterminacy
on the value of $\eta_{2}$ (and thus of $\mu_{c}$) is strongly related
to the fact that we cannot generate decreasing curves within the relaxed
micromorphic model. This issue will be solved in a subsequent work.
At the present stage, we chose to fix the value of the micro-rotation
inertia $\eta_{2}$ to be the same of the micro-distortion inertia
$\eta_{1}$ (which actually produces the desired horizontal $TO_{2}$
curve) and we thus compute the corresponding value of $\mu_{c}$.
Except for this indeterminacy on the parameters $\eta_{2}$ and $\mu_{c}$,
all the remaining parameters of the model are uniquely calibrated.
\item The parameter $\eta_{3}$ is seen to have a non-negligible effect
on the optic curve $LO_{2}$ and on the longitudinal acoustic curve
$LA$ and is hence adjusted in order to obtain the best possible fitting.
On the other hand, the parameter $\bar{\eta}_{2}$ is seen to have
the opposite effect on the curve $LO_{2}$ as well as a slight effect
on the curve $TA$. The parameters $\eta_{3}$ and $\bar{\eta}_{2}$
are then carefully calibrated in order to get the best possible agreement
of the curve $LO_{2}$ without macroscopically perturbing the other
curves affected by such parameters.
\item The found set of values for $\eta_{1}$, $\eta_{2}$, $\eta_{3}$,
$\bar{\eta}_{1}$, $\bar{\eta}_{2}$, $\bar{\eta}_{3}$, is then slightly
arranged for a last refined fitting. The obtained numerical values
are shown in Table \eqref{tab:fitted}. The remaining parameters of
the model can also be determined by means of the relations \eqref{eq:partial_Solution}
and are shown in Table \eqref{tab:fitted_2}.
\end{itemize}
\begin{table}[H]
\begin{centering}
\begin{tabular}{|c|c|c|c|c|c|}
\hline 
$\eta_{1}$  & $\eta_{2}$  & $\eta_{3}$  & $\bar{\eta}_{1}$  & $\bar{\eta}_{2}$  & $\bar{\eta}_{3}$ \tabularnewline
\hline 
$\left[\mathrm{kg/m}\right]$ & $\left[\mathrm{kg/m}\right]$ & $\left[\mathrm{kg/m}\right]$ & $\left[\mathrm{kg/m}\right]$ & $\left[\mathrm{kg/m}\right]$ & $\left[\mathrm{kg/m}\right]$\tabularnewline
\hline 
\hline 
$3.25\times10^{-5}$ & $3.25\times10^{-5}$ & $4\times10^{-4}$ & $0$ & $0.3\times10^{-4}$ & $1.8\times10^{-4}$\tabularnewline
\hline 
\end{tabular}
\par\end{centering}
\caption{\label{tab:fitted}Values of the micro-inertia parameters of the weighted
relaxed micromorphic model as fitted on the metamaterial shown in
Figure \ref{fig:micro_1}.}
\end{table}

\begin{table}
\begin{centering}
\begin{tabular}{|c|c|c|c|c|c|c|}
\hline 
$\rho$  & $\mu_{c}$ & $\lambda_{\mathrm{micro}}$ & $\mu_{\mathrm{micro}}$ & $\lambda_{e}$ & $\mu_{e}$ & $L_{c}$ \tabularnewline
\hline 
$\left[\mathrm{kg/m^{3}}\right]$ & $\left[\mathrm{GPa}\right]$ & $\left[\mathrm{GPa}\right]$ & $\left[\mathrm{GPa}\right]$ & $\left[\mathrm{GPa}\right]$ & $\left[\mathrm{GPa}\right]$ & $\left[\mathrm{m}\right]$\tabularnewline
\hline 
\hline 
$1323$ & $0.272$ & $19.8$ & $0.737$ & $17.7$ & $3.857$ & $0$\tabularnewline
\hline 
\end{tabular}
\par\end{centering}
\caption{\label{tab:fitted_2}Values of the parameters of the weighted relaxed
micromorphic model for the metamaterial shown in Figure \ref{fig:micro_1}.}
\end{table}

\begin{figure}[H]
\begin{centering}
\begin{tabular}{cc}
\negthickspace{}\negthickspace{}\negthickspace{}\negthickspace{}\negthickspace{}\negthickspace{}\negthickspace{}\negthickspace{}\negthickspace{}\negthickspace{}\negthickspace{}\negthickspace{}\negthickspace{}\negthickspace{}\negthickspace{}\negthickspace{}\negthickspace{}\includegraphics[scale=0.35]{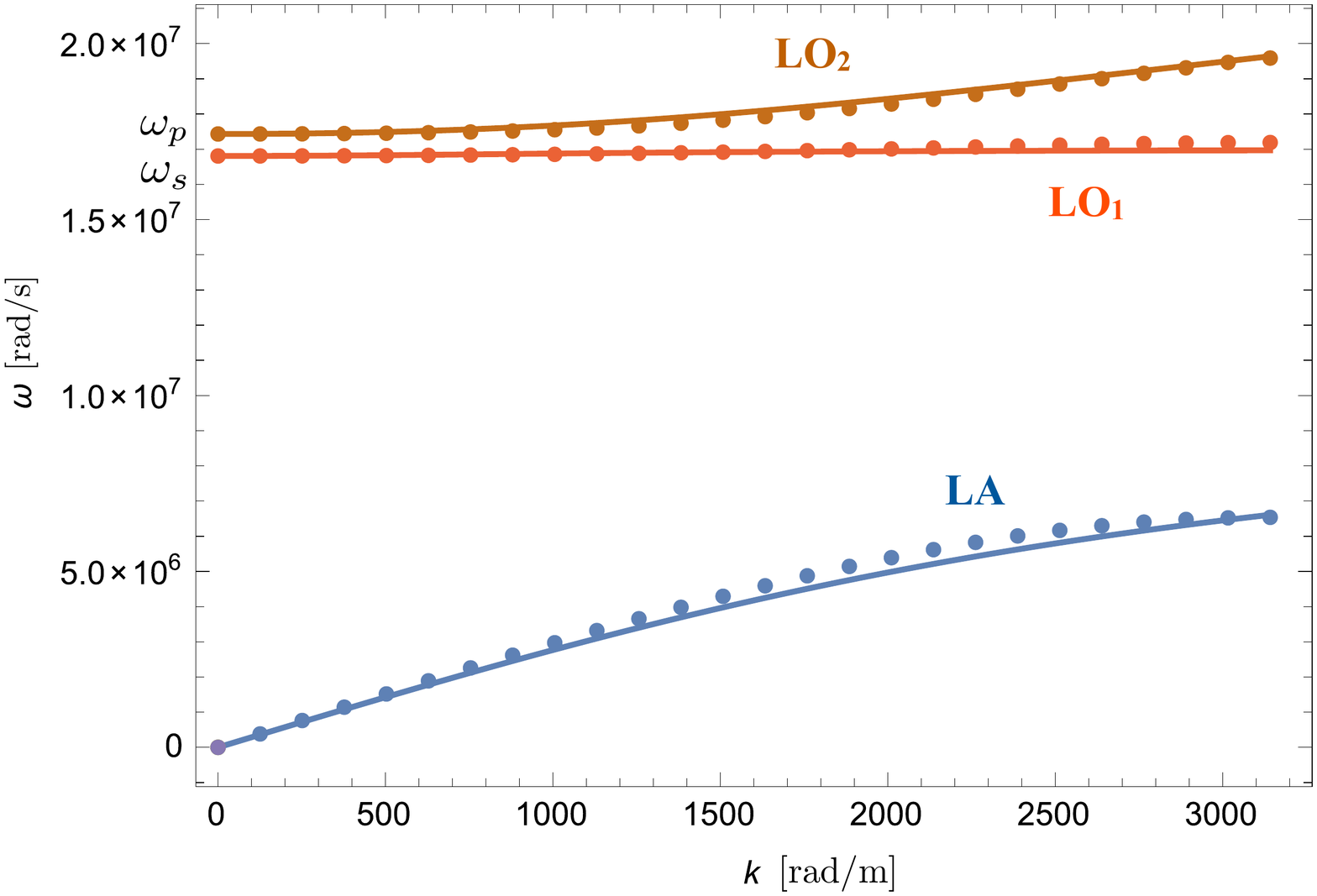} & \negthickspace{}\negthickspace{}\negthickspace{}\negthickspace{}\negthickspace{}\negthickspace{}\negthickspace{}\negthickspace{}\negthickspace{}\negthickspace{}\negthickspace{}\negthickspace{}\negthickspace{}\negthickspace{}\negthickspace{}\negthickspace{}\negthickspace{}\negthickspace{}\includegraphics[scale=0.35]{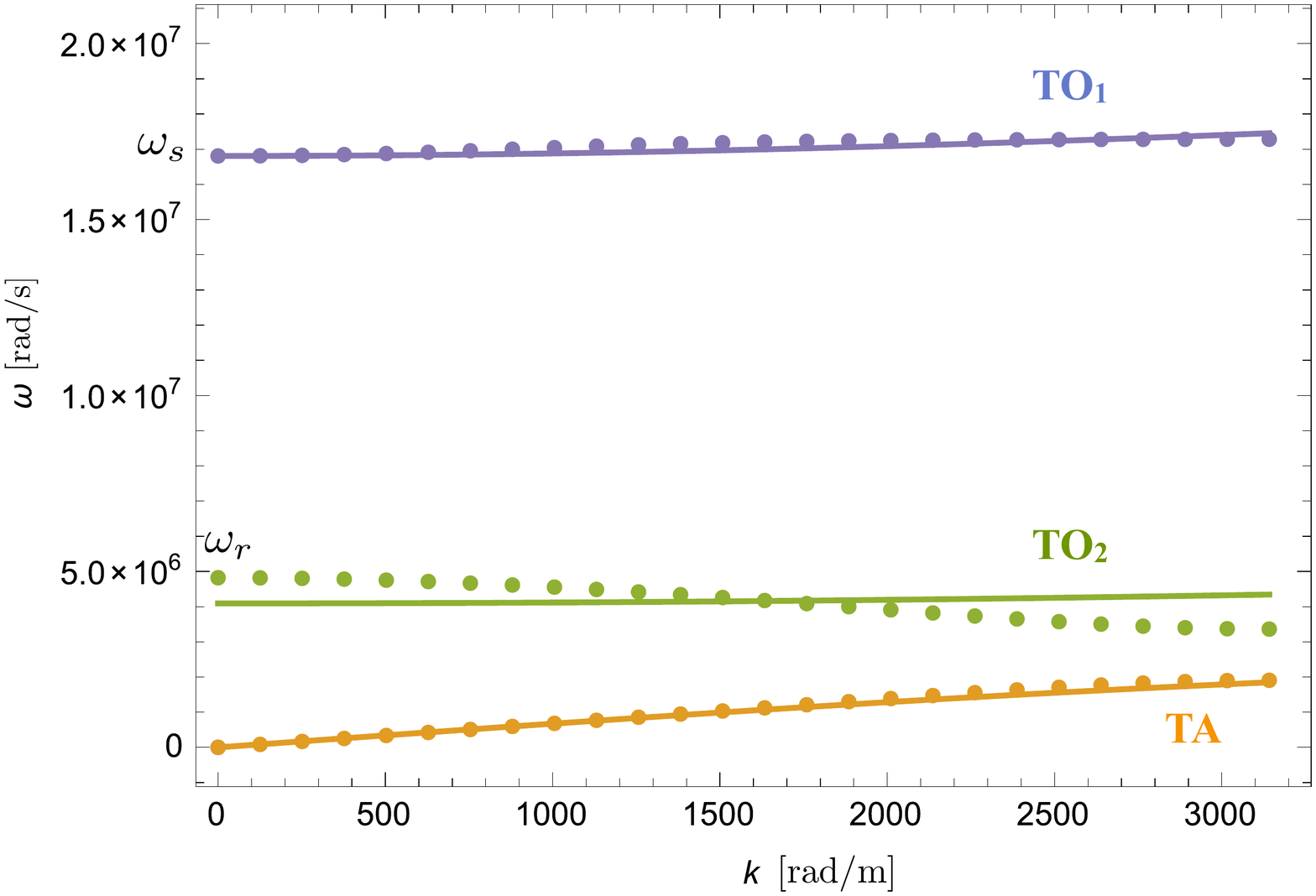}\tabularnewline
\end{tabular}
\par\end{centering}
\centering{}\caption{\label{fig:Final_Fitting}Fitting of the curves obtained via the weighted
relaxed micromorphic model (continuous lines) on the dispersion curves
issued via the Bloch wave analysis (dotted curves), for wavenumbers
up to the size of the unitary cell. Left: longitudinal waves; Right:
transverse waves.}
\end{figure}

The final fitting of the parameters of the relaxed micromorphic model
on the curves issued by the Bloch wave analysis is shown in Figure
\eqref{fig:Final_Fitting}. An almost perfect fitting is obtained
for all the curves, with the exception of some small deviations of
the $TO_{2}$ curve which cannot at the present stage be arranged
to be a decreasing curve. This last detail will be fixed in subsequent
works.

According to the fitted values of the parameters and considering formulas
\eqref{eq:Micro_Macro}, it is also possible to derive, \textit{a
posteriori}, the value of the averaged macroscopic parameters of the
considered metamaterial, which are shown in Table \ref{tab:macroCoeff}.
\begin{table}[H]
\begin{centering}
\begin{tabular}{|c|c|}
\hline 
$\lambda_{\mathrm{macro}}$ & $\mu_{\mathrm{macro}}$\tabularnewline
\hline 
$\left[\mathrm{GPa}\right]$ & $\left[\mathrm{GPa}\right]$\tabularnewline
\hline 
\hline 
$0.62$ & $4.66$\tabularnewline
\hline 
\end{tabular}\qquad{}\qquad{}\qquad{}%
\begin{tabular}{|c|c|}
\hline 
$E_{\mathrm{macro}}$ & $\nu_{\mathrm{macro}}$\tabularnewline
\hline 
$\left[\mathrm{GPa}\right]$ & $\left[-\right]$\tabularnewline
\hline 
\hline 
$1.78$ & $0.44$\tabularnewline
\hline 
\end{tabular}
\par\end{centering}
\caption{\label{tab:macroCoeff}Values of the macroscopic parameters of the
considered metamaterial computed starting from the values shown in
Tables \ref{tab:fitted} and \ref{tab:fitted_2} and using the homogenization
formulas \eqref{eq:Micro_Macro} (left) and corresponding values of
the Young modulus and Poisson ratio (right).}
\end{table}
 It is possible to notice, comparing the values of the parameters
of the metamaterial given in Table \ref{tab:macroCoeff} with those
of the original material (aluminum) shown in Table \ref{fig:micro_1},
that the presence of the cross cavity renders the resulting metamaterial
softer than the original one. Also the Poisson's effect results to
be enhanced in the metamaterial compared to the original material. 

\newpage

\subsection{Analysis of the vibrational modes}

In this subsection we analyze the vibrational modes of the considered
metamaterial as function of the frequency and of the wavelength, both
theoretically via the relaxed micromorphic model and by using the
Bloch wave analysis.

\begin{figure}[H]
\begin{centering}
\includegraphics[scale=0.4]{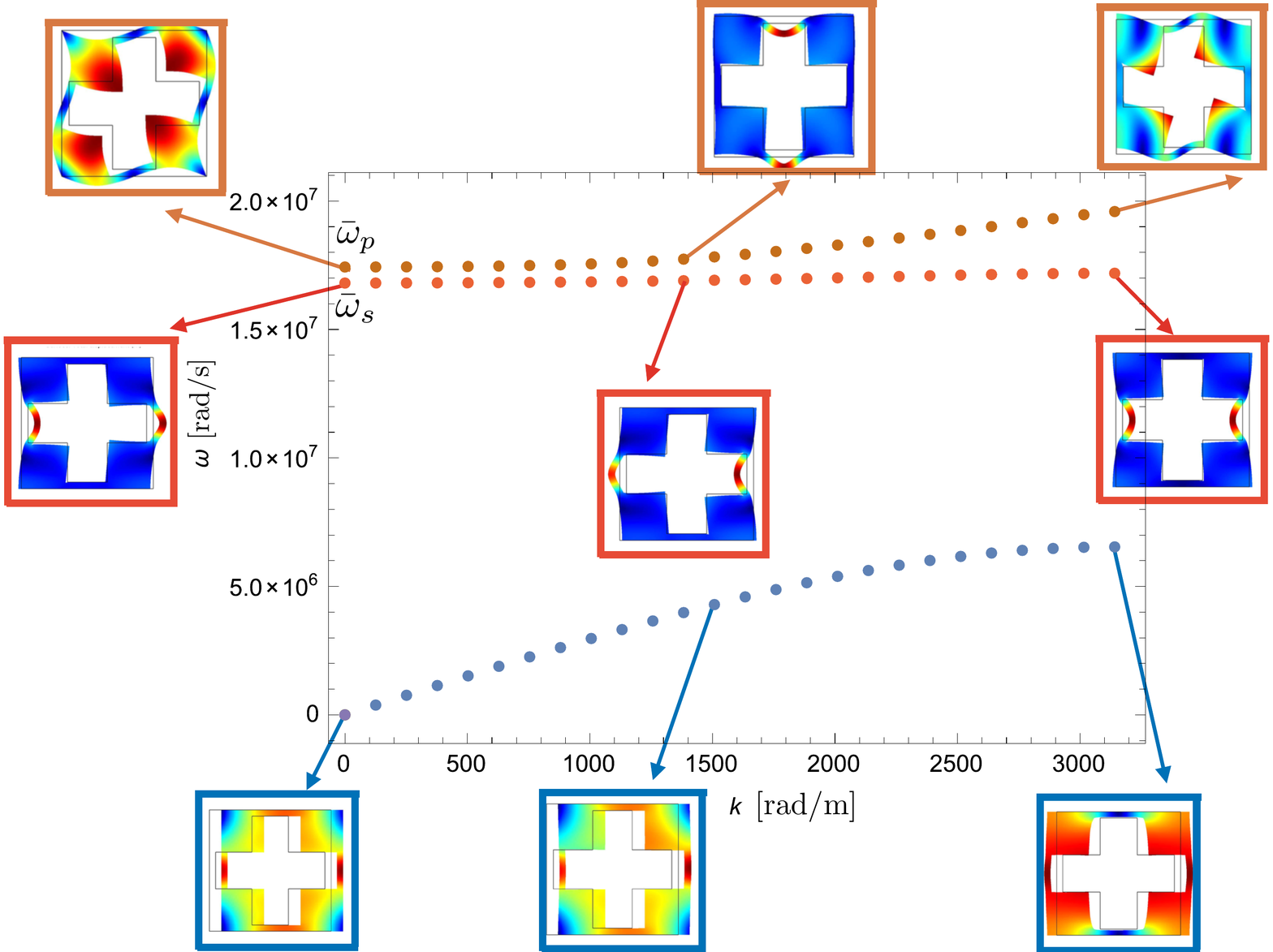}
\par\end{centering}
\caption{\label{fig:modi} Vibrational modes for longitudinal waves as function
of the wavenumber obtained via Bloch wave analysis. }
\end{figure}
\begin{figure}[H]
\begin{centering}
\includegraphics[scale=0.4]{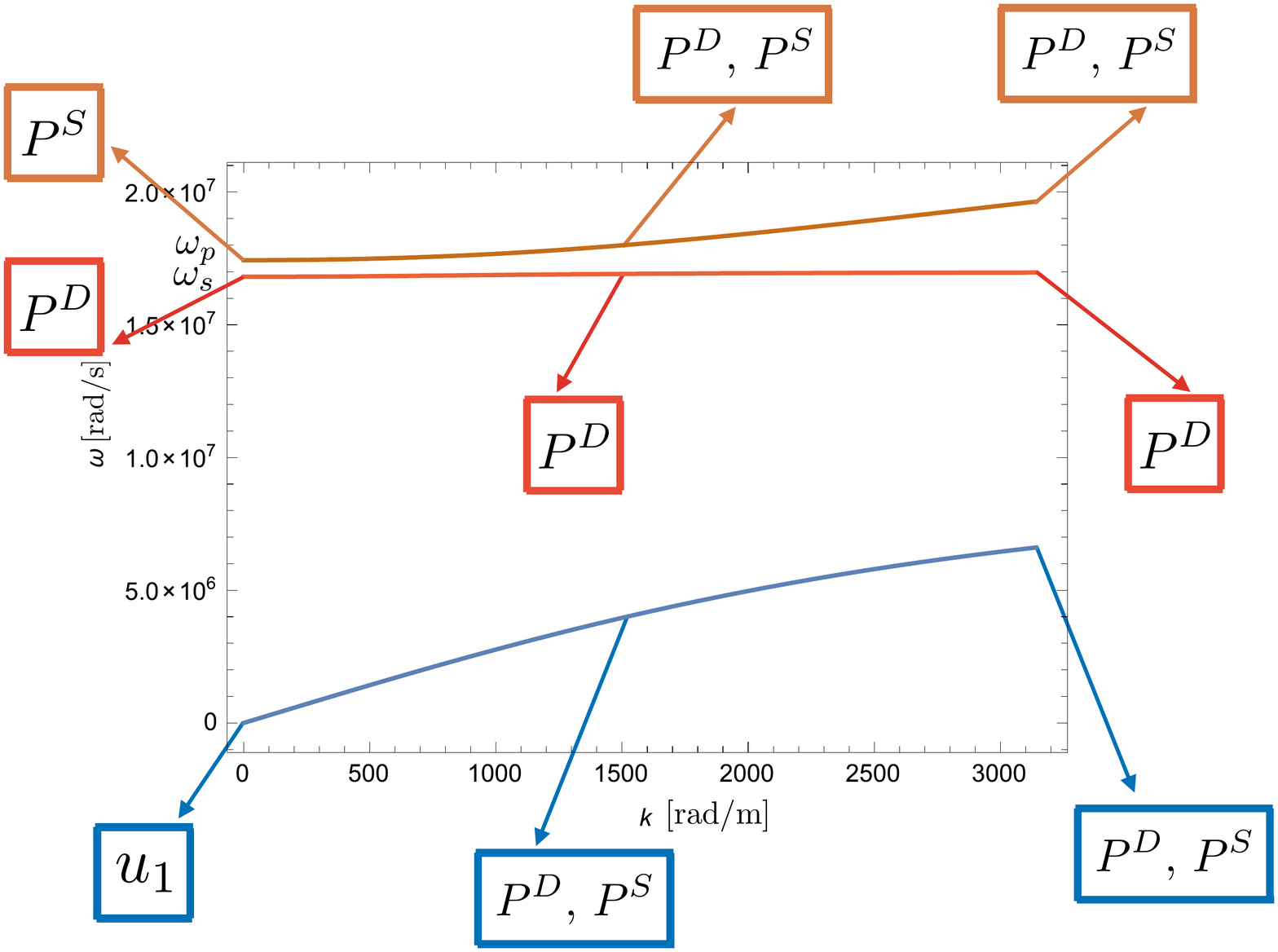}
\par\end{centering}
\caption{\label{fig:modi_teorici} Qualitative distribution of the theoretical
vibration modes for longitudinal waves obtained via the relaxed micromorphic
model as function of the wavenumber. }
\end{figure}

\begin{figure}[H]
\begin{centering}
\includegraphics[scale=0.4]{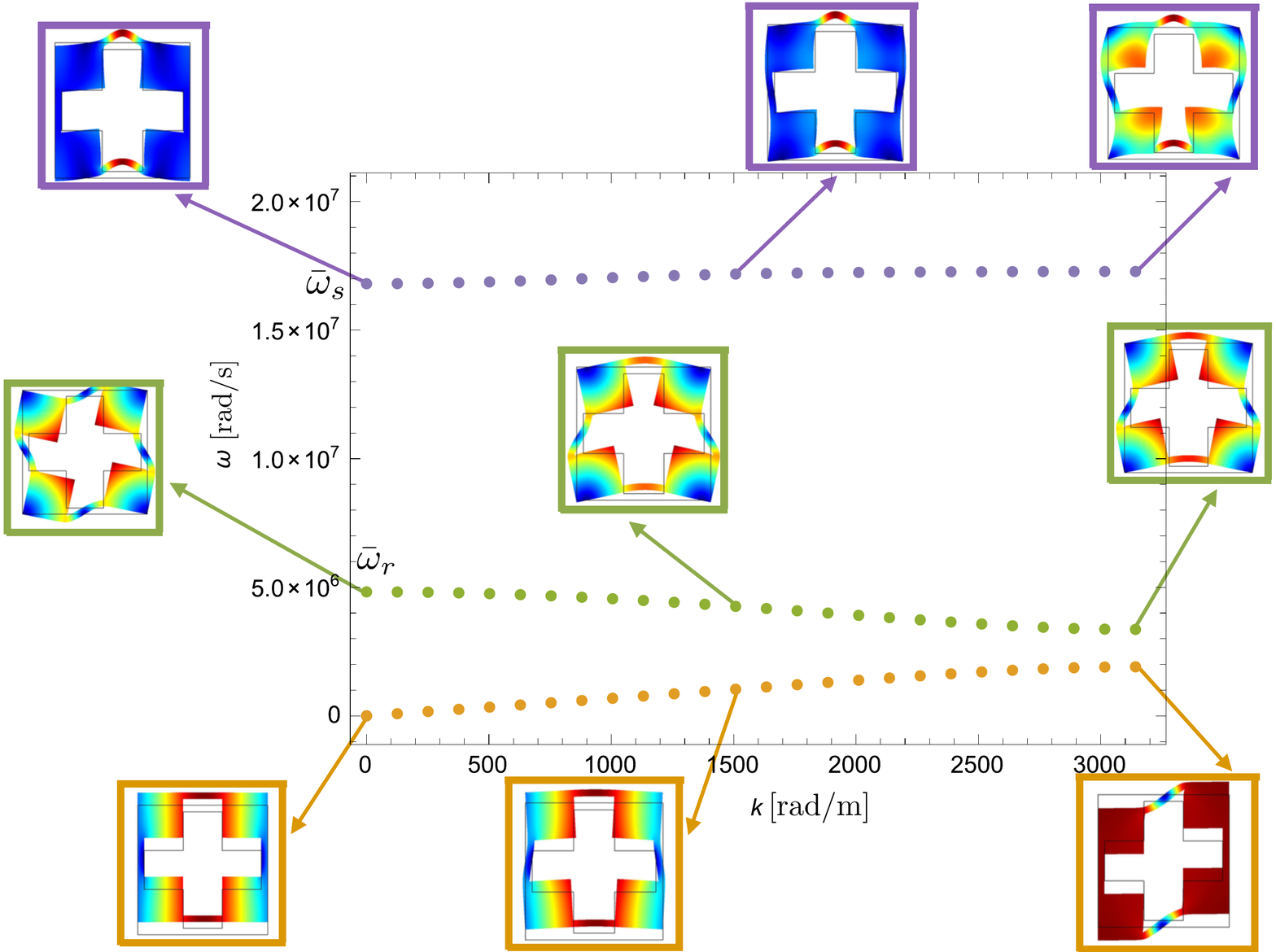}
\par\end{centering}
\caption{\label{fig:modi-1} Vibrational modes for transverse waves as function
of the wavenumber obtained via Bloch wave analysis. }
\end{figure}
\begin{figure}[H]
\begin{centering}
\includegraphics[scale=0.4]{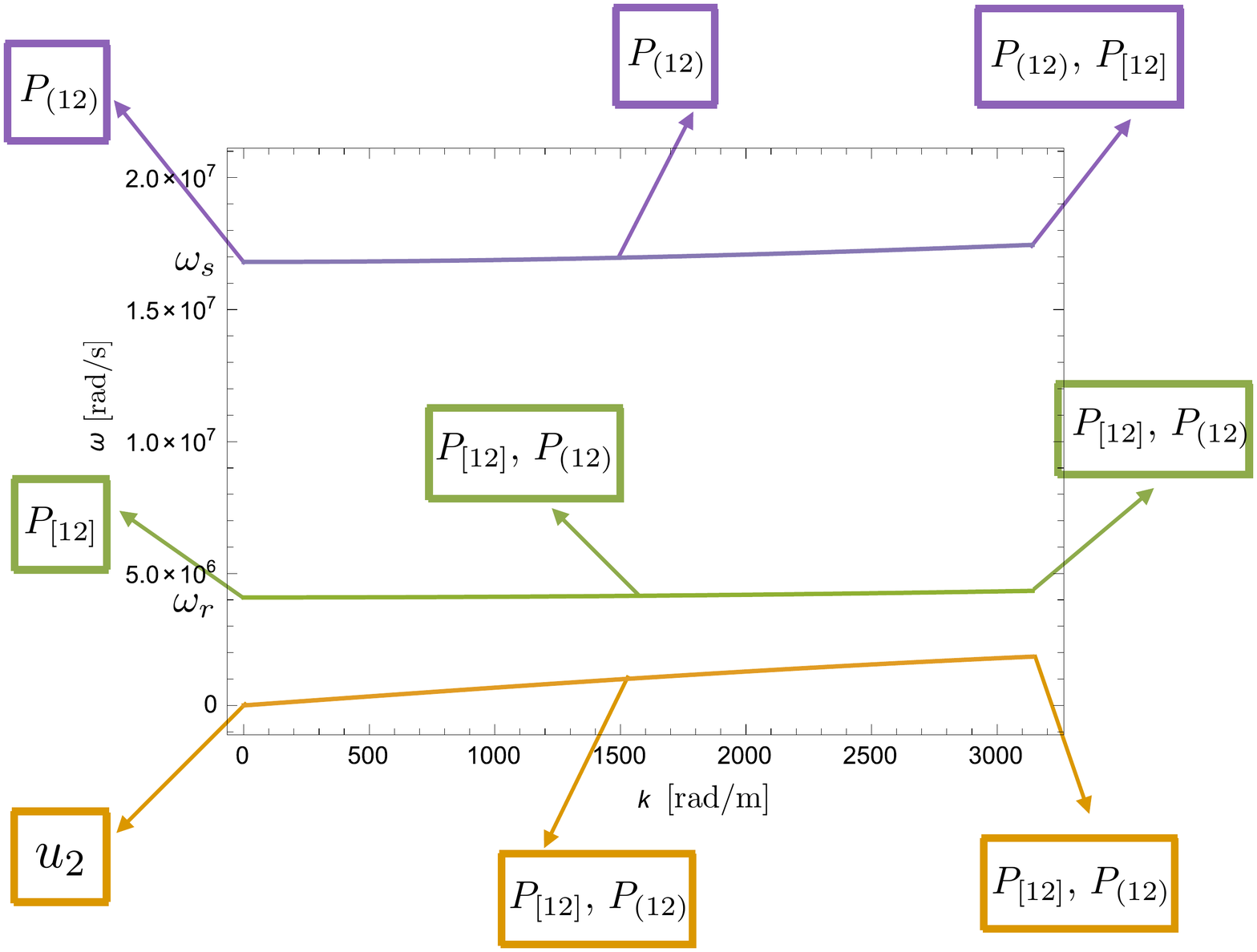}
\par\end{centering}
\caption{\label{fig:modi_teorici-1} Qualitative distribution of the theoretical
vibration modes for transverse waves obtained via the relaxed micromorphic
model as function of the wavenumber. }
\end{figure}

Indeed, it is well known that the vibrational modes of the unit cell
of periodic metamaterials vary when varying the wavelength of the
traveling wave. This can be recognized in Figure \ref{fig:modi} where
the main vibrational modes of the unit cell are depicted as function
of the wavenumber up to the size of the unit cell. It can be recognized
that for the longitudinal and transverse acoustic waves, the main
vibrational modes are given by horizontal and transverse macroscopic
displacements of the unit cell when considering very large wavelength
(small $k$). When increasing the wavenumber (reducing the wavelength)
we can observe from Figures \ref{fig:modi} and \ref{fig:modi-1}
that for $k\sim1500\,\mathrm{rad/m}^{-1}$ the microstructure-related
acoustic vibrational modes start to take a predominant role. This
means that the unit cell is not only displaced of a given amount,
but also starts deforming. Volume variations coupled to micro-distortion
of the unit cell can be observed for the longitudinal modes, while
micro-rotations coupled to in-plane shear are found for the transverse
modes. Analogous patterns can be obtained via the relaxed micromorphic
model when normalizing the eigenvectors corresponding to the eigenvalues
shown in Figure \ref{fig:Final_Fitting}. In particular, the agreement
of the vibrational modes found via the relaxed micromorphic model
with those issued via the Bloch analysis can be deciphered when recalling
that 
\begin{itemize}
\item $P^{S}$ is the spherical part of the micro-distortion tensor related
to volume variations, 
\item $P^{D}$ is related to the deviatoric part of the micro-distortion
tensor which is known to be associated to the actual distortion of
the unit cell
\item $P_{(12)}$ is the $(1,2)$ component of the symmetric part of the
micro-distortion tensor which can be directly related to in-plane
shear of the unit cell,
\item $P_{[12]}$ is the $(1,2)$ component of the skew-symmetric part of
the micro-distortion tensor which is known to be related to in-plane
rotations of the microstruture embedded in the unit cell.
\end{itemize}

\section{Conclusions}

In this paper we study the macroscopic behavior of real band-gap metamaterials
by using the linearized, isotropic relaxed micromorphic model with
weighted free and gradient micro-inertia. For a specific microstructure,
we make a direct comparison between the dispersion curves issued via
the classical Bloch wave analysis and those obtained by means of our
weighted relaxed micromorphic model. Such comparison allows us to
uniquely identify almost all the parameters of the model with the
exception of the characteristic length $L_{c}$ and of the Cosserat
couple modulus $\mu_{c}$. In particular, the characteristic length
$L_{c}$ is a very sensitive parameter measuring the non-locality
of the considered metamaterial, the determination of which requires
a much more refined fitting procedure than that which is possible
on the basis of the dispersion curves alone. A more refined fitting
based e.g. on the energy which is transmitted through the considered
metamaterial for different frequencies is needed, as done e.g. in
\cite{madeo2016first}. The determination of the characteristic length
of the metamaterial targeted in this paper will be then analyzed in
subsequent works, given that the measure of all the remaining parameters
is not affected by the determination of $L_{c}$.

On the other hand, the Cosserat couple modulus $\mu_{c}$ is found
to be directly proportional to the micro-inertia parameter $\eta_{2}$
(related to the skew-symmetric part of the tensor $P_{,t}$). Nevertheless,
the performed fitting does not allow to uniquely conclude about the
value of $\mu_{c}$. Choosing the rotational micro-inertia $\eta_{2}$
to be equal to the distortion micro-inertia $\eta_{1}$, we are able
to deduce the corresponding value of $\mu_{c}$. To allow a more precise
fitting of $\mu_{c}$, the relaxed micromorphic model needs to be
further generalized in order to grant the possibility of dispersion
curves which are decreasing for increasing wavenumber $k$. Such a
generalization of the relaxed micromorphic model will be the focus
of subsequent works.

The results presented in this paper represent a topical breakthrough
allowing a simple implementation of the relaxed micromorphic model
in finite element codes in view of meta-structural design.

\section{Acknowledgments}

Angela Madeo thanks INSA-Lyon for the funding of the BQR 2016 \textquotedbl{}Caractérisation
mécanique inverse des métamatériaux: modélisation, identification
expérimentale des paramètres et évolutions possibles\textquotedbl{},
as well as the CNRS-INSIS for the funding of the PEPS project.

{\footnotesize{}\let\stdsection\section \def\section *#1{\stdsection{#1}}}{\footnotesize \par}

\bibliographystyle{plain}
\bibliography{library}

\end{document}